\documentstyle[twocolumn,psfig,rotating]{mn2e}

\def\lsim{~\rlap{$<$}{\lower 1.0ex\hbox{$\sim$}}}
\def\bsim{~\rlap{$>$}{\lower 1.0ex\hbox{$\sim$}}}
\def\kms{\ {\rm km\,s^{-1}}}
\def\skm{\ {\rm s\,km^{-1}}}

\def\hmpc{\ {\rm {\it h}^{-1}Mpc}}

\def\hmmpc{\ {\rm {\it h}Mpc^{-1}}}
\def\cmm{\ {\rm cm^{-2}}}

\def\kel{\,{\rm K}}

\def\la{\langle}
\def\ra{\rangle}
\def\dd{{\rm d}}
\def\ln{{\rm ln}}

\def\vv{{\bf v}}
\def\vp{{\bf p}}

\def\vnabla{{\bf\nabla}}

\def\xf{x_{\rm F}}
\def\kf{k_{\rm F}}

\def\db{\delta_{\rm b}}
\def\vb{v_{\rm b}}
\def\dl{\delta_{\rm L}}
\def\vl{v_{\rm L}}

\def\sgl{\sigma_{\rm L}}

\def\anu{\alpha_\nu}
\def\ala{\alpha_\lambda}
\def\wb{W_{\rm b}}

\def\dhi{n_{_{\rm HI}}}
\def\dhhi{{\hat n}_{_{\rm HI}}}
\def\nhi{N_{_{\rm HI}}}

\def\thg{\hat{T}_{\rm g}}
\def\thgg{\hat{T}_4}

\def\dlk{\Delta_{\rm L}^2(k)}

\def\dlw{\Delta_{\rm w}^2}
\def\dlu{\Delta_{\rm u}^2}
\def\djj{\Delta_{\rm 3D}^2}
\def\dii{\Delta_{\rm 1D}^2}

\def\tef{\tau_{\rm eff}}

\def\pmb#1{\setbox0=\hbox{#1}%
\kern-.025em\copy0\kern-\wd0
\kern.05em\copy0\kern-\wd0
\kern-.025em\raise.0433em\box0}

\def\etal{{\it et al.\ }}

\newcommand{\op}{Ly$\alpha$\ }
\newcommand{\hi}{\mbox{H{\scriptsize I}}}

\newcommand{\heii}{\mbox{He{\scriptsize II}}}
\newcommand{\oxi}{\mbox{O{\scriptsize I}}}
\newcommand{\nai}{\mbox{Na{\scriptsize I}}}

\begin{document}

\title[] {The probability distribution of the \op transmitted flux 
from a sample of SDSS quasars}

\author[Desjacques {\it et al.}]{Vincent Desjacques$^{1,2}$, Adi Nusser$^2$ 
and Ravi K. Sheth$^3$ \\
$^1$ Racah Institute of Physics, The Hebrew University, Jerusalem 91904, 
Israel \\
$^2$ Physics Department, The Technion, Haifa 32000, Israel \\
$^3$ Department of Physics \& Astronomy, University of Pennsylvania, 
Philadelphia, PA 19104, USA \\
Email~: dvince@phys.huji.ac.il\\}

\maketitle

\begin{abstract}
We present a measurement of the probability distribution function
(PDF) of the transmitted flux in the \op forest from a sample of
3492 quasars included in the SDSS DR3 data release.  Our intention is 
to investigate the sensitivity of the \op flux PDF  as measured from low
resolution and low signal-to-noise data to a number  of systematic errors
such as uncertainties in the mean flux, continuum  and noise
estimate. The quasar continuum  is described by the superposition of a
power law and emission  lines. 
We perform a power law continuum fitting on a spectrum-by-spectrum basis,
and obtain an average  continuum slope of
$\anu=0.59\pm 0.36$ in the redshift range  $2.5<z<3.5$. Taking into
account the variation in the continuum indices increases the mean flux 
by 3 and 7 per cent at $z=3$ and 2.4, respectively, as compared to the 
values inferred with a single (mean) continuum slope.  We compare our
measurements to the PDF obtained with mock lognormal spectra,  whose
statistical properties have been constrained to match the  observed
\op flux PDF and power spectrum of high resolution data. Using our
power law continuum fitting and the SDSS pipeline noise estimate yields
a poor agreement between the observed and mock PDFs. Allowing for a 
break in the continuum slope and, more importantly, for residual scatter 
in the continuum level substantially improves the agreement. 
A decrease of $\sim$10-15 per cent in the mean 
quasar continuum with a typical rms variance at the 20 per cent level 
can account for the data, provided that the noise excess correction is 
no larger than $\lsim 10$ per cent.
\end{abstract}


\section{Introduction}
\label{sec:intro}

The \op forest seen in quasar spectra probes the intergalactic medium
(IGM) and the underlying matter distribution over a wide range of
scales ($k\sim 0.1-10\hmmpc$) and redshifts ($1\lsim z\lsim 6$).
Measurements of the mean flux in the \op  forest shed light  on the
reionization history and the physical state of the IGM in  the
post-reionization area (Press, \& Rybicki \& Schneider 1993, hereafter
P93; Rauch \etal 1997; Bernardi \etal 2003, hereafter B03; Bolton \&
Haehnelt 2006). Fluctuations in the \op flux are of great  interest
since they provide information on the matter distribution on  scales
smaller than those accessible to other observables (e.g. Croft \etal
1998; 1999; 2002b; Nusser \& Haehnelt 1999; 2000; Pichon \etal 2002;
Zaldarriaga, Hui \& Tegmark 2003; McDonald \etal 2000; McDonald \etal 
2005a). Combined with CMB observations, the power spectrum of the \op 
forest can provide stringent constraints  on the shape and amplitude 
of the primordial power spectrum (Seljak  \etal 2004; Viel \& Haehnelt 
2005).

The probability distribution function (PDF) of the \op transmitted 
flux was first
studied by Jenkins \& Ostriker (1991). It has, however, received less
attention in the past years as it is more sensitive to systematics
errors such as continuum fitting uncertainties. 
Yet, the tension between the WMAP and \op values of the
normalisation amplitude $\sigma_8$ argues in favour of incorporating 
statistics others than the \op flux power spectrum. 
Rauch \etal 1997 and
McDonald \etal (2000) have computed the PDF of the \op transmitted flux
from  high resolution data and found that $\Lambda$CDM cosmologies
provide a  good fit to the observations. Gazta\~{n}aga \& Croft (1999)
have provided an analytic description of  the \op flux PDF based on
perturbation theory. Choudhury,  Srianand \& Padmanabhan (2001), and
Desjacques \& Nusser (2005)  have investigated the  constraints obtained
from a joint  fit of the PDF  and power spectrum of the \op
transmitted flux. Becker, Rauch \& Sargent (2006) have examined the 
redshift evolution of the flux \op PDF from a large sample of Keck 
HIRES data. Note also that Lidz \etal (2005) have advocated working
with the PDF of the fluctuations in the flux about the mean as it is
insensitive to the quasar continuum. 

The Sloan Digital Sky Survey (SDSS; York \etal 2000) has greatly
increased the statistical power of the \op forest. Unfortunately,
attempts to exploit the data are plagued by a number of poorly
constrained parameters that describe the physical state of the IGM, by
inaccuracies in the numerical modelling of the \op forest, and by
systematics errors in the measurement such as continuum fitting
uncertainties.  The observed flux in the \op region of a quasar (QSO)
spectrum depends both on the quasar continuum (the flux emitted by the
quasar, including the emission lines) and on the amount of absorption
by intervening  galactic matter (lines and continuum absorptions). The
transmitted (or normalized) flux  obtained by continuum fitting of the 
observed spectrum is related to  the optical depth along the line of 
sight as
\begin{equation}                                                            
F(w)\equiv I_{\rm obs}(w)/I_{\rm cont}={\rm e}^{-\tau(w)}\; ,
\label{eq:cont}
\end{equation}
where $\tau$ is the optical depth, $w$ is the redshift space
coordinate along the line of sight, $I_{\rm obs}$ is the observed flux
and $I_{\rm cont}$ is the flux emitted from the source (quasar) that
would be observed in the absence of any intervening material.  To
estimate the unabsorbed continuum, two different approaches have been
used. When the signal-to-noise is large, a polynomial continuum  is
fitted to regions of the \op forest free of absorption (e.g. Rauch
\etal 1997; McDonald \etal 2000). When the signal-to-noise is low, one
usually extrapolate the continuum redward of the \op emission  line,
assuming a power law shape. The last method only provides an
approximation to the true continuum, and can easily introduce several
types of systematic errors (e.g. Steidel \& Sargent 1987).  The strong
degeneracy between the effective optical depth  $\tef=-\ln\la F\ra$
and the continuum level affects the determination of the clustering
amplitude $\sigma_8$.  Measurements of the effective optical depth
from low resolution quasar spectra with comparable signal-to-noise
(P93; B03) are systematically higher than those based on
high-resolution  spectra by 10-20 per cent. As argued by Seljak  \etal
(2003), Tytler  \etal (2004), Viel \etal (2004b), this is most  likely
due to systematic  errors in the continuum fitting procedure of the
low resolution spectra.  Furthermore, the low signal-to-noise of SDSS
spectra in the \op region ($S/N\sim 3$ typically) requires an accurate
noise estimate. Systematic errors in the noise characterisation will
affect the accuracy of the measurements. In this respect, several
lines of recent evidence suggest that the SDSS reduction pipeline
underestimates the true noise by 5-10 per cent (McDonald \etal 2006;
Burgess 2004).
 
In this paper, we measure the PDF  of the \op transmitted flux from 
a large (public) sample of quasars included in the SDSS DR3 data
release (Abazajian \etal 2005). 
We assume that the quasar continuum follows the parametric form
given in B03. However, unlike B03, we also allow for
the variation in the continuum indices of individual spectra and fit 
a power law continuum for each spectrum separately. 
We compare our measurements to the
probability distribution obtained from lognormal, realistic looking
SDSS spectra. The  statistical properties of these mock spectra are
beforehand constrained to match the observed \op flux probability 
distribution and power  spectrum of high resolution data of the forest. 
Our intention is to investigate the sensitivity of the \op flux PDF 
as measured from low resolution, low signal-to-noise data to a number 
of systematic errors such as uncertainties in the mean flux, continuum 
and noise estimate. 

The paper is organised as follows. We briefly review the lognormal
model of the \op forest in Section~\S\ref{sec:lya}. The constraints on
the model parameters are discussed in~\S\ref{sec:param}. The continuum
fitting procedure and the measurement of the PDF of the  \op flux are
presented in~\S\ref{sec:data}. In~\S\ref{sec:constrain}, we compare
the simulated and observed PDF, and study the effect of a number of
systematic errors. We discuss our results  in~\S\ref{sec:discussion}. 
In ~\S\ref{sec:conclusion}, we conclude and indicate potential future 
works. We will present results  for
a $\Lambda$CDM cosmology with normalisation amplitude $\sigma_8=0.83$,
and spectral index $n_s=0.96$. This is consistent with the constraints
obtained from the latest CMB and \op forest data (Spergel \etal 2006;
Viel, Haehnelt \&  Lewis 2006; Seljak, Slosar \& McDonald 2006).

\section{Generating mock quasar spectra}
\label{sec:lya}

We implement the lognormal model introduced by Bi and collaborators
(Bi, B\"orner \& Chu 1992; Bi 1993; Bi \& Davidsen 1997; see also
Choudhury, Padmanabhan \& Srianand 2001; Viel \etal 2002) to simulate
the distribution of low-column density \op absorption lines along the
LOS to quasars. 
The main advantage of this procedure is that simulated spectra 
can have an arbitrary large length. 
This allows us to eliminate periodicity effects that are present in
simulations, where the typical box size is noticeably  smaller than
the total length of a single spectrum.
Furthermore, this approach is computationally very efficient as
compared to N-body simulations.

\subsection{The lognormal model of the \op forest}
\label{sub:log}

The lognormal model of the IGM is based on the assumption that the
low-column density \op forest is produced by mildly nonlinear 
fluctuations ($\delta\rho/\rho\lsim 10$) which smoothly trace the 
dark matter distribution.
The IGM density contrast $\db$ is obtained from a local mapping of 
the linear IGM density contrast $\dl$ (Coles \& Jones 1991), but 
the IGM peculiar velocity along the line of sight is assumed to be 
linear even on scales where the 
density contrast gets non-linear (Bi \& Davidsen 1997)~\footnote{
This assumption is motivated by the continuity equation which reads
$\vnabla\cdot\vv\propto -\dd\ln(1+\delta)/\dd t$ if one neglects
the coupling $\delta\vv$.}. We have namely
\begin{eqnarray}
\db(x,z) &=& \exp\left(\dl(x,z)-\sgl^2(z)/2\right)-1 \nonumber \\
\vb(x,z) &=& \vl(x,z) \;,
\label{eq:lg}
\end{eqnarray}
where $\sgl(z)$ is the rms fluctuations of the linear IGM density 
field at redshift $z$.
The linear IGM density and peculiar velocity, $\dl$ and $\vl$, are 
obtained by smoothing the linear matter density and velocity 
distribution on some characteristic scale $\xf=1/\kf$ to mimic
pressure smoothing. The linear IGM clustering amplitude, $\sgl$, is
thus 
\begin{equation}
\sgl^2=D(z)^2\int_0^\infty\!\!\dd\ln k\,\dlk \wb^2(k,\kf)\;,
\label{eq:rmsigm}
\end{equation}
where $D(z)$ is the linear density growth factor, $\dlk$ is the 
dimensionless, linear matter power spectrum at the present epoch 
(Peebles 1980) and $\wb(k,\kf)$ is the IGM filter. We 
take the $k$-space filter $\wb$ to be a Gaussian. Such a filter 
gives a good fit to the gas fluctuations over a wide range of wavenumber 
(Gnedin \etal 2003; see also Zaroubi \etal 2005).
In principle, we expect $\kf$ to depend on the physical state of the
IGM. However, since the relation between $\kf$ and $\thg$ depends
noticeably on the reionization history of the Universe (Gnedin \& Hui
1998; Nusser 2000), it is more  convenient to treat $\kf$ as a free 
parameter. 

We calculate the \op transmitted flux $F=\exp(-\tau)$ in the fluctuating 
Gunn-Peterson approximation (Gunn \& Peterson 1965, Bahcall \& Salpeter 
1965). The optical depth for \op resonant scattering at some redshift 
space position $\omega$ is expressed as a convolution of the real space 
\hi\ density along the line of sight with a Voigt profile ${\cal H}$,
\begin{equation}
\tau(w)=\frac{c \sigma_0}{H(z)}\int_{-\infty}^{+\infty}\!\! dx\,
\dhi(x){\cal H} \left[w-x-\vb(x),b(x)\right]\; .
\label{eq:depth}
\end{equation}
where $\sigma_0=4.45\times 10^{-18}\cmm$ is the effective
cross-section for resonant line scattering, H$(z)$ is the Hubble
constant at redshift $z$, $x$ is the real space coordinate, $\dhi(x)$ 
is the neutral hydrogen density, $b(x)$ is the Doppler parameter due 
to thermal/turbulent broadening and ${\cal H}(x)$ is the Voigt profile, 
which can be approximated by a Gaussian for moderate optical depths.
The \hi\ hydrogen density, $\dhi(x)$, and the Doppler parameter, $b(x)$, 
are computed using a tight polytropic relation (Katz \etal 1996, Hui 
\& Gnedin 1997, Theuns \etal 1998) that produces results 
comparable to full hydrodynamical simulations,
\begin{eqnarray}
\dhi(x) &=& \dhhi\left(1+\db\right)^{2-0.7(\gamma-1)} \nonumber \\
b(x) &=& \hat{b}\left(1+\db\right)^{(\gamma-1)/2}\;,
\label{eq:hi}
\end{eqnarray}
where the adiabatic index $\gamma$ is in the range $1-1.6$. $\dhhi$ and 
$\hat{b}$ are the \hi\ hydrogen density and Doppler parameter at mean gas 
density respectively. The latter is a function of the IGM temperature, 
$\hat{b}=13\kms\thgg^{1/2}$, where $\thgg$ is the IGM temperature at 
mean density (in unit of $10^4\kel$). Note that, since $\dhhi$ is 
usually constrained by fixing the mean flux level, $\la F\ra$, we 
will treat $\la F\ra$ as a free parameter in the remaining of this 
paper.

\subsection{The simulation method}
\label{sub:los}

We simulate the \op transmitted flux in a periodic line of sight of 
comoving length $L$ at $N$ discrete $r$-space position $x_i$, 
$i=1,\cdots,N$. Following Bi (1993), we generate two fields $\dl(k)$ 
and $\vl(k)$ at discrete $k$-space positions. The two fields are 
correlated Gaussian random fields that can be written as linear 
combination of two independent fields $u(k)$ and $w(k)$ having 
(dimensionless) power spectra $\dlu(k)$ and $\dlw(k)$,
\begin{eqnarray}
\dl(k,z) &=& D(z)\left(u(k)+w(k)\right) \nonumber \\
\vl(k,z) &=& iqE(z)\frac{H_0}{c}f(k)w(k)\,,
\label{eq:uw}
\end{eqnarray}
where $E(z)$ is the linear growth factor of the velocity field. 
$\dlu(k)$, $\dlw(k)$ and $f(k)$ are constrained by the auto- and 
cross-correlations of the density and velocity fields,
\begin{eqnarray}
f(k) &=& \int_k^\infty\!\!\dd\ln q\, q^{-5}\djj(q)\,/
\int_k^\infty\!\!\dd\ln q\, q^{-3}\djj(q) \nonumber \\
\dlw(k) &=& f(k)^{-1}\,\int_k^\infty\!\!\dd\ln q\, q^{-3}\djj(q) 
\nonumber \\
\dlu(k) &=& \dii(k)-\dlw(k)\;,  
\label{eq:uwf}
\end{eqnarray}
where $\djj$ and $\dii$ are the 3D and 1D power spectra, respectively.
We obtain the linear IGM density and peculiar velocity field in $r$-space 
from a fast Fourier transform (FFT), and calculate the mean transmitted flux
by combining equations (\ref{eq:lg}), (\ref{eq:depth}) and (\ref{eq:hi}). 
For a given cosmological model, the flux distribution depends only on the 
filtering wavenumber $\kf$, the mean flux $\la F\ra$,  the adiabatic index 
$\gamma$ and the mean IGM temperature $\thgg$.

\subsection{Properties of the synthetic spectra}
\label{sub:syn}

To create realistic mock spectra, we compute the flux distribution on 
a one-dimensional (1D) grid whose resolution $\Delta$ is fine enough to 
resolve the smallest structure on the filtering scale, and whose length 
$L$ is long enough to incorporate most of the fluctuation power. We 
typically have $\Delta\lsim 0.002\hmpc$ and $L\bsim 1000\hmpc$ (comoving).
We constrain the mean flux level $\la F\ra$ from a large sample of 
idealised, noise-free spectra. Instrumental resolution, noise and 
strong absorption systems are included as described below. Note that, 
in low resolution spectra, $\la F\ra$ will generally differ from the 
``effective'' mean flux $\bar{F}$ of the processed spectra, which 
include a number of strong absorption systems. 

\subsubsection{Instrumental noise and resolution}

We attempt to include instrumental noise and resolution in our
synthetic spectra in a way that mimics the  observations as much as
possible (e.g. Rauch \etal 1997). We smooth the spectra  with a
Gaussian of constant width, and re-sample them  on pixels  of
wavelength size $\Delta\lambda$,  interpolating between adjacent
values in the fine grid.  The spectra are not modified to include
continuum fitting uncertainties as the latter are taken out from the
observed spectra.

In the high resolution spectra, the transmitted flux is convolved with
a Gaussian of full width at half maximum (FWHM) $6.6\kms$, and
re-sampled on pixels of wavelength size $\Delta\lambda=0.04\AA$. In
units of $\kms$, the resolution varies from $2\kms$ at $z=3.9$
to $2.9\kms$ at $z=2.4$. Further, Gaussian noise is added to each
pixel with an amplitude given by a flux-dependent rms noise per pixel,  
$n(F)$, as measured in M00 (see their Table 3).

In the low resolution spectra, we adopt a FWHM of $170\kms$ (i.e. a
dispersion of $\approx 70\kms$), and re-sample the flux on a uniform
grid with $\Delta\log_{10}\lambda=10^{-4}$.  The low
signal-to-noise of the SDSS spectra in the \op  region requires an
accurate description of the noise.  The true error on each pixel
in each quasar spectrum is essentially made  up of the Poisson noise
from photon counts, the CCD read-noise and  systematics errors from
sky subtraction. McDonald \etal (2006) have outlined a procedure 
which follows closely the spectroscopic data reduction pipeline. The
inconvenience of this method is that it requires a sky-flux estimate.
Here, we simply assume that the noise distribution is Gaussian
whose rms variance is given by the SDSS data reduction pipeline.
Although the latter computes  a variance for each  flux pixel, there
is some evidence that these error estimates do not  perfectly reflect
the true errors in the data (Bolton \etal 2004; McDonald \etal 2006;
Burgess 2004).  McDonald \etal (2006) and Burgess (2004) have
recalibrated the noise  of their spectra by differencing multiple
exposures, and found that  the rms noise variance given by the SDSS
pipeline is underestimated  by resp. 8 and 5 per cent on average. We 
will hereafter refer to $\sigma_p$ as the fiducial (pipeline) noise
estimate. The sensitivity of the mock spectra to the noise level will 
be discussed in detail in~\S\ref{sub:sens}. Given the relatively broad 
distribution of pixel noise variances at fixed flux $F$, we use the
full set of rms noise variances from the data points with $0<F<1$. 
We then map with repetition the individual noise estimates onto the
pixels $0<F<1$ in the idealized mock spectra (e.g. Burgess 2004).
Our mapping accounts for the fact that the mean pixel noise increases
with redshift (see Figure~\ref{fig:pix}).

\begin{figure}
\resizebox{0.48\textwidth}{!}{\includegraphics{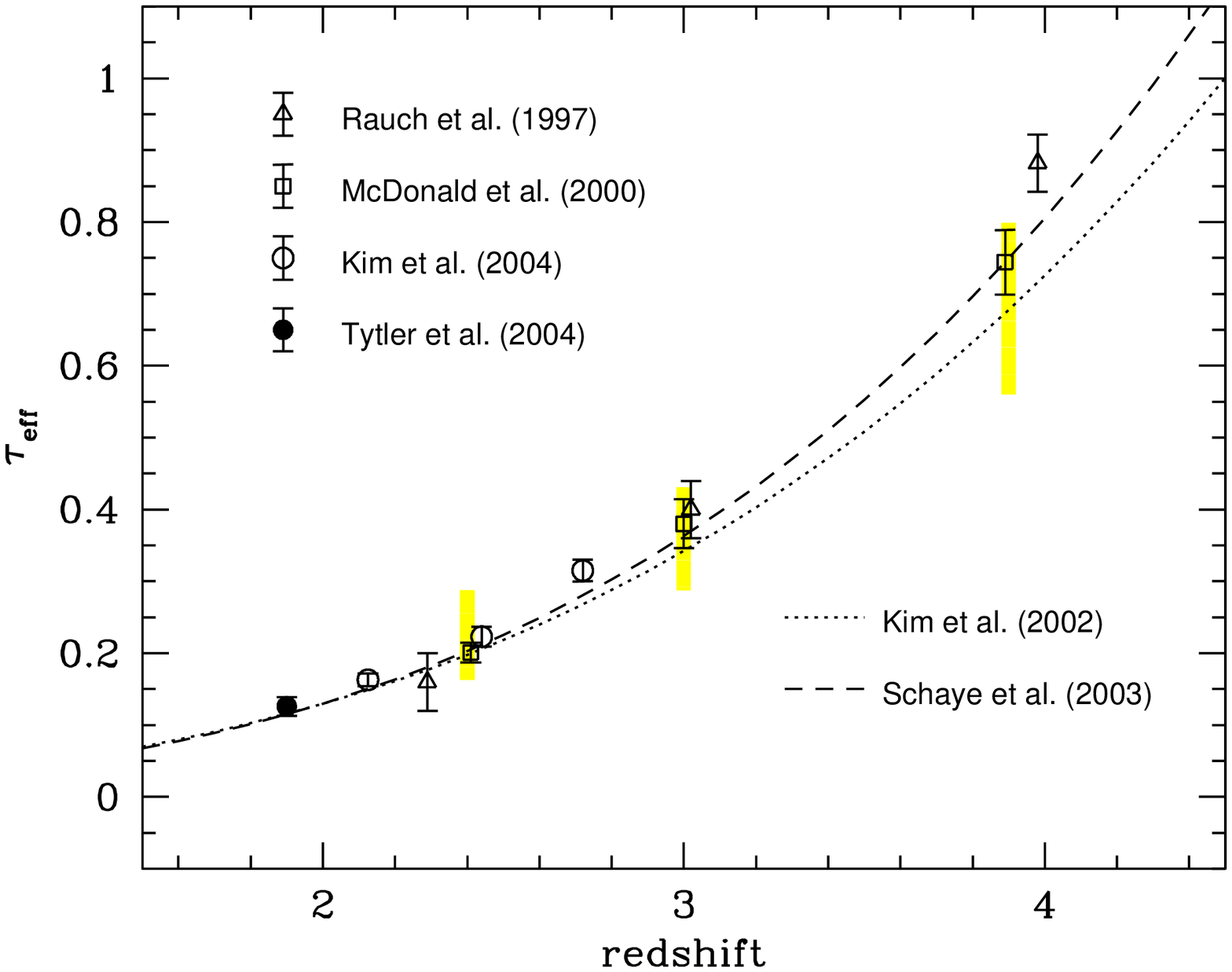}}
\caption{The effective optical depth $\tef=-\ln\la F\ra$ as measured
from high-resolution spectra. The empty triangles, squares  and
circles show the measurements of Rauch \etal (1997), McDonald \etal
(2000) and Kim \etal (2004). The filled circle is the measurement of
Tytler \etal (2004). The dotted and dashed curves show the evolution
reported  by Kim \etal (2002) and Schaye \etal (2003). The vertical
extent of the shaded regions  indicate the values considered in our
analysis.}
\label{fig:mteff}
\end{figure}

\subsubsection{Strong absorption systems}

High column density systems associated with  collapsed objects such as
disk galaxies are not reproduced in the  lognormal model (Bi \&
Davidsen 1997).  Large column densities are needed ($\nhi\bsim
10^{17}\cmm$) to produce  the strong damping wings of the observed
absorption line profile.  In the high-resolution data, damped \op
systems (DLAs) are removed by eliminating wavelength intervals
containing these absorption lines from the spectra. These strong
absorption lines are, however,  present in our sample of SDSS spectra.
We, therefore, include a set of DLAs and Lyman limit systems (LSS) 
in the low resolution mock spectra. We follow the  procedure
outlined in McDonald \etal (2005b), which is based on the
self-shielding model of Zheng \& Miralda-Escud\'e (2002). The column
density distribution of these strong absorbers is then normalised to
reproduce the observations of Peroux \etal 2003 and Prochaska,
Herbert-Fort \& Wolfe (2005). As the Doppler parameter distribution
of the strong absorption systems is largely unknown, we use 
a single value of 30$\kms$ at all redshifts.

\section{Constraining the model parameters from high resolution data}
\label{sec:param}

In this Section, we constrain the model parameters from a comparison
between the flux power spectrum (PS) and probability distribution
(PDF) of the transmitted flux of mock and observed high resolution
spectra.   Desjacques \& Nusser (2005) have pointed out that models
that match best the PS alone do not necessarily yield a good fit to
the PDF.  It is therefore important to combine the PS and PDF
statistics to  ensure that both are correctly reproduced in the
lognormal spectra.  We perform a $\chi^2$ statistical test for the
observed flux power  spectrum and PDF to determine quantitatively the
values of the parameter required to fit high resolution measurements
of the \op forest.

\subsection{The high resolution data}

\begin{figure}
\resizebox{0.48\textwidth}{!}{\includegraphics{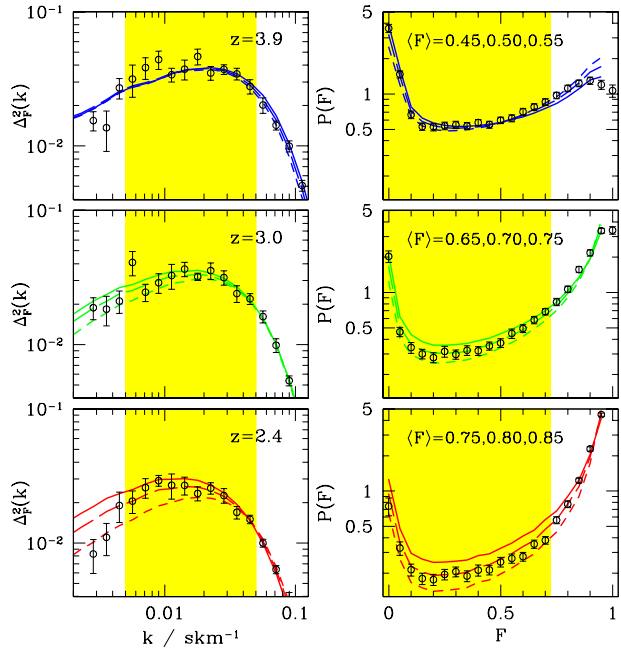}}
\caption{The flux power spectrum and PDF of the best-fitting
models ($\Delta\chi^2=0$) at redshift $z=3.9$ (top), 3.0 (middle) and
2.4 (bottom). The mean flux is, respectively, $\la F\ra=0.45$, 0.5, 0.55,
$\la F\ra=0.65$, 0.7, 0.75, and $\la F\ra=0.75$, 0.8, 0.85. In each 
panel, the solid and short-dashed curves stand for the lowest and 
largest values of $\la F\ra$, respectively. The IGM temperature has a 
fixed value, $\thgg=1.5$. Only $\kf$ and $\gamma$ are varied to obtain 
the best-fitting parameters. The shaded regions indicate the data points
we use to compute the value of $\chi^2$. The best-fitting values of $\kf$ 
and $\gamma$, together with the chi-squared value are listed in Table 
~\ref{table:best}.}
\label{fig:best}
\end{figure}

We use the measurements of McDonald \etal (2000), which were obtained
from a sample of eight high resolution QSO spectra. Results are
provided  for three redshift bins centered at $z=2.41$, $3.00$ and
$3.89$.  Regarding the flux power spectrum, we consider the data
points in the  range $0.005<k<0.05\skm$. The lower limit $k=0.005\skm$
is chosen so as to avoid continuum fitting errors (Hui \etal 2001),
and the upper limit $k=0.05\skm$ is chosen to avoid
metal contamination  on smaller scales (Kim \etal 2004).   The
observed flux PDF is very sensitive to continuum fitting, especially
in the high transmissivity tail (e.g. Meiksin, Bryan \& Machacek
2001). The modelling of these errors is complicated by the fact
that the scales of interest are of the order of the box size $L$ of
the simulations.  However, M00 demonstrate that, if the
inclusion of continuum fitting errors can account for most of the
discrepancy between the simulated and observed PDF in the range
$F\bsim 0.8$, it should not greatly affect the PDF for $F\lsim 0.8$. 
We, therefore, exclude the data points   with $F\geq 0.8$ from
the analysis to avoid dealing with those errors. The M00 measurements 
are shown in Figure~\ref{fig:best} as the filled symbols. The shaded 
areas indicate the data points used in this analysis is (10+15=25 
measurements from the PS and PDF respectively).

\subsection{The parameter grid}

For each value of the parameter vector 
$\vp$=($\kf$,$\la F\ra$,$\gamma$,$\thgg$), 
we generate mock catalogues of 1000 lines of sight. We let the filtering 
wavenumber, the IGM adiabatic index and temperature assume the following 
values, 
\begin{eqnarray*}
\kf &=& 5.55,6.25,7.14,8.33,10,12.5,16.67,25,50 \\ 
\gamma &=& 1,1.2,1.4,1.6 \\ 
\thgg &=& 1,1.5,2,2.5, 
\end{eqnarray*}
irrespective of the redshift. The values of $\kf$ (in unit of $\hmmpc$) 
are chosen such that $1/\kf$  uniformly spans the range $0.02-0.18\hmpc$.  

The assumed effective optical depth $\tef$ or, equivalently, mean
flux $\la F\ra=\exp(-\tef)$ has  a large impact on the simulated one-
and two-point statistics of the forest. Observations indicate that
$\tef$ evolves strongly in the redshift range $2\lsim z\lsim 4$. In
Fig.\ref{fig:mteff}, we show several measurements of $\tef$ obtained
from high resolution observations. Empty triangles, squares and
circles show the results of Rauch \etal (1997) and McDonald \etal
(2000) for a comparable sample of HIRES spectra, whereas the empty
circles indicate the estimates of the LUQAS sample of Kim \etal
(2004).  The filled circle shows the measurements of Tytler \etal
(2004). The  dotted and dashed curves indicate the evolution reported
by Kim \etal (2002) and Schaye \etal (2003). Note that there is 
significant overlap among the quasar samples of Schaye \etal (2003)
and Kim \etal (2004). All these estimates have been obtained after
removing damped/sub-damped \op systems and pixels contaminated by
associated metal absorption.  The measurements of $\tef$ are mostly
affected by cosmic variance due to large variations between lines of
sight, uncertainties in the continuum fitting procedure and the
somewhat uncertain  contribution from metal lines (P93;  Zuo \& Bond
1994; Rauch \etal 1997;  Tytler \etal 2004; Viel \etal 2004a).  In
particular, the continuum fitting generally adopted for high
resolution data may result in an underestimation of the continuum and
of $\tef$ (Kim \etal 2001). Based on these measurements, we  adopt the
following values for the mean flux $\la F\ra$~:   $0.45\leq\la
F\ra\leq 0.55$ ($z=2.4$), $0.65\leq\la F\ra\leq 0.75$  ($z=3$) and
$0.75\leq\la F\ra\leq 0.85$ ($z=3.9$). These intervals are shown as
shaded regions in  Fig.~\ref{fig:mteff}. Although the effective
optical depth appears to evolve smoothly with redshift, the behaviour
of $\tef(z)$ inferred from a large sample of SDSS quasars is found to
deviate from a power law around $z=3.2$ (B03), suggesting that \heii\
reionizes in that redshift range (Theuns \etal  2002b; Schaye \etal
2000).

We use a spectral grid of $N=2^{19}$ pixels which are evenly spaced in 
wavelength ($\Delta\lambda=0.005\AA$). $N$ and $\Delta$ are
chosen so that the number of pixels in the ``degraded'' mock spectra 
is a constant power of 2 to facilitate the computation of Fourier 
transforms. 

For each mock catalogue, we calculate the flux power spectrum and the  
flux PDF. We determine the goodness  of fit of any model in the grid by
computing a $\chi^2$ statistic from the difference between the simulated 
PS and PDF and the observational data.  
In the calculation of the $\chi^2$, we neglect the correlations between
measurements of the flux PS. However, in the case of the flux PDF, we 
include the full covariance matrix since these measuremements are highly 
correlated (M00).
We take advantage of the smooth dependence of the flux PS and PDF on 
the parameter vector, and use cubic spline interpolation to find the
best-fitting models. 

\begin{table}
\caption{Parameter values of the models which best fit the PS and PDF 
inferred from high-resolution measurements of the \op forest. The 
mean IGM temperature has a fixed value $\thgg=1.5$. Only $\kf$ and
$\gamma$ are varied to obtain the best-fitting models. The filtering 
$\kf$ is in unit of $\hmmpc$. The last column gives the chi-squared for 
23 degrees of freedom. Note that, since we spline interpolate over the 
parameters, the best fit values do not necessarily lie at a grid point.}
\vspace{1mm}
\begin{center}
\begin{tabular}{cc|ccc|c} \hline
redshift & $\la F\ra$ & $\kf$ & $\gamma$ & $\chi^2$ \\ \hline
3.9 & 0.45 & 37.1 & 1.00 & 24.6 \\
    & 0.50 & 32.7 & 1.00 & 26.7 \\
    & 0.55 & 36.3 & 1.12 & 45.7 \\
3.0 & 0.65 & 18.3 & 1.00 & 39.5 \\
    & 0.70 & 20.5 & 1.00 & 22.6 \\
    & 0.75 & 32.8 & 1.16 & 40.7 \\
2.4 & 0.75 & 12.5 & 1.00 & 108.8 \\
    & 0.80 & 15.5 & 1.00 & 41.8 \\
    & 0.85 & 25.0 & 1.00 & 46.3 \\
\hline\hline
\end{tabular}
\end{center}
\label{table:best}
\end{table}

\subsection{The best-fitting models}

Fig.~\ref{fig:best} compares the best-fitting models to the M00 data at
redshift $z=3.9$ (top panels), 3.0 (middle panels) and 2.4 (bottom
panels).  The parameter values of the models are listed in
Table~\ref{table:best}.  The solid, long dashed and short dashed
curves show, respectively, the models with lowest, intermediate and  
largest value of
$\la F\ra$ at a given redshift. The IGM temperature has a  fixed
value, $\thgg=1.5$. Only $\kf$ and $\gamma$ are varied to obtain  the
best-fitting parameters. The shaded regions indicate the data points
we use to compute the value of $\chi^2$.  Note that the 1D grid
resolves the best-fitting values of the filtering length with 10 cells
at least.  The best-fitting models provide an acceptable fit to the
data down to  redshift $z=3$. At $z=3.9$ and 3, the best chi-squared
has  an acceptable value of $\chi^2\lsim 25$ for 23 degrees  of
freedom (25  data points minus the filtering length and adiabatic
index). At $z=2.4$ however, $\chi^2\bsim 40$, which should be exceeded
randomly only  $\sim$1.5 per cent of the time. As expected, the
lognormal approximation  no longer provides a good fit (in a
chi-squared sense at least) to the data for $z<3$ (e.g. Nusser \&
Haehnelt 2000). For most of the models,  the best-fitting value of the
adiabatic index is $\gamma=1$, whereas observations indicate that
$\gamma\sim 1.3-1.5$ in the redshift range considered here  (Schaye
\etal 2000b; McDonald \& Miralda-Escud\'e 2001).   Note, however,
that there is a degeneracy between the filtering  wavenumber $\kf$ and
the adiabatic index $\gamma$ which allows one to match the data with
larger values of $\gamma$ and $\kf$ (Desjacques \&  Nusser 2005; see
also Meiksin \& White 2001). 

The M00 results are averaged in relatively large  redshift bins,
$2.09\leq z\leq 2.67$, $2.67\leq z\leq 3.39$ and $3.39\leq z\leq
4.43$. The evolution of the mean flux $\la F\ra$, for example, is
significant over those redshift intervals. We could account for the
redshift evolution by averaging mock catalogues computed at  different
$z$ in the same redshift bin before comparing with the  observations.
However, this correction is difficult to apply as the exact dependence
of $\kf$, $\gamma$ and $\thgg$ on the redshift is unknown.  Additional
assumptions on the reionization history of the Universe could reduce
the freedom in the parameter space.  In this respect, the observed
line-width distribution suggests that, around $z=3$, there is a sharp
increase in $\thg$ together with a decrease  in $\gamma$ (Schaye \etal
2000b; Ricotti, Gnedin \& Shull  2000; McDonald \& Miralda-Escud\'e
2001).  However, the data are too  noisy to provide robust constraints 
on $\thgg$ and $\gamma$.

\section{The data}
\label{sec:data}

\subsection{The SDSS DR3 sample}
\label{sub:data}

\begin{figure}
\resizebox{0.45\textwidth}{!}{\includegraphics{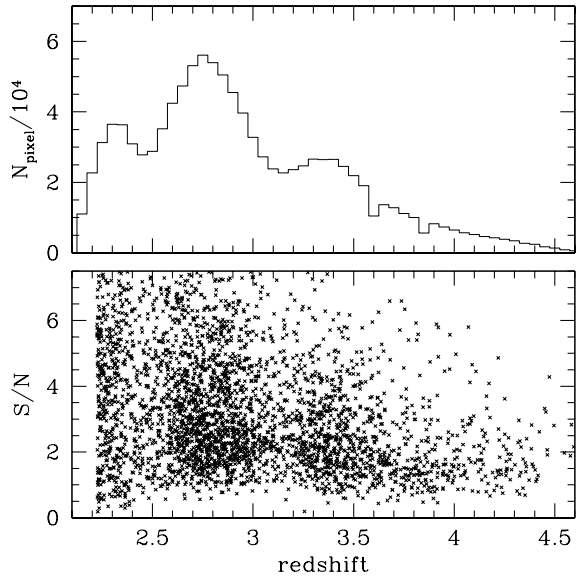}}
\caption{
{\it Top}~: Distribution of the \op forest pixels in the data as
a function of redshift. There are typically $350$ pixels per spectrum 
which lie in the \op forest. The gaps at $z=3.59$ and $3.84$ correspond 
to the OI and NaI lines. 
{\it Bottom}~: Average signal-to-noise ratios in the \op forest as a 
function of its median redshift for the quasars in the DR3 sample.} 
\label{fig:pix}
\end{figure}

We use 3492 quasar spectra included in the Sloan Digital Sky
Survey DR3 data release
(Abazajian \etal 2005). York \etal (2000) provide a technical
summary of the survey. The SDSS camera and the filter response
curves are described in Gunn \etal (1998) and Fukugita \etal
(1996), respectively. Lupton \etal (2001) and Hogg \etal 
(2001) discuss the SDSS photometric data and monitoring system.
Richards \etal (2002) describe the algorithm for targeting 
quasar candidates from the multi-color imaging SDSS data.
To avoid contamination from Ly$\beta$ absorption and the proximity 
effect on the blue and red sides of the \op forest, we define the
\op forest as the rest-frame interval $1080-1160\AA$ (B03).
The top panel of Fig.~\ref{fig:pix} shows the number of pixels in 
the DR3 sample which belong to the \op forest as defined above. 
The gaps at $z\simeq 3.59$ and $z\simeq 3.84$ correspond to the
\oxi (5577\AA) skyline, and interstellar line \nai (5894.6\AA) 
respectively. Pixels in the wavelength range 
$5570\leq\lambda\leq 5590$ and $5885\leq\lambda\leq 5905\AA$ 
were removed from the analysis. 
The fainter quasars, mostly those at redshift $z>4$, suffer from 
significant contamination from OH emission features. Although 
these OH sky-subtraction residuals could in principle be removed 
(e.g. Wild \& Hewett 2005), we have not bothered to do so because 
they mostly affect pixels longward of 6700\AA.

The distribution of the signal-to-noise ratios in the \op forest  as a
function of the median redshift $z_{\rm Ly\alpha}$ is plotted  in the
bottom panel of Fig.~\ref{fig:pix}. The transmitted flux in the forest 
is lower at high-redshifts, so higher redshift  spectra tend to
have lower signal-to-noise ratios. The typical  signal-to-noise ratio
in the \op forest is $S/N\simeq 4.6$, 3.8 and 2.1 at 
redshift $z=2.4$, 3.0 and 3.9, respectively.  As a result, most of the 
\op absorption lines are unresolved in  the data.

\subsection{Estimating the continuum}
\label{sub:cont}

In high resolution and high signal-to-noise spectra, the shape of the
continuum is determined separately for each QSO. A polynomial
continuum is fitted to regions of the \op forest which are free of
absorption lines (as judged by eye). In low resolution observations
such as the SDSS-DR3 sample, an object-by-object estimate of the
continuum is difficult. However, the large size of the sample is
suitable for a statistical approach. This allowed B03 to
constrain simultaneously the mean quasar continuum and mean
transmitted flux in the \op forest.  There is indeed a remarkable
similarity between the spectra of the most distant quasars at $z\bsim
6$ and their low redshift counterparts (Fan \etal 2003). At fixed
luminosity, the spectral properties of quasars show  little evolution
with cosmic epoch (Vanden Berk \etal 2004 ).  Hence, the mean QSO
continuum can be thought of as being representative of the  quasar
population as a whole.

The mean continuum is usually calibrated redward  of the \op emission
line and then extrapolated blueward assuming a smooth power law shape
(P93).  Composite spectra suggest that the shape of the quasar
continuum is  the superposition of a single power law and emission
lines. A principal component analysis (PCA) demonstrates that this is a
reasonable assumption redward of  the \op emission line, where QSOs
differ significantly in the normalisation, but little in the shape of
the continuum (e.g. Yip \etal 2004).  It is unclear, however, whether 
this parametrisation can be extended to  wavelengths blueward of the \op
emission line given the large  impact of intervening absorption.  At
low redshift, where absorption in the \op range is much less
significant, composite spectra of IUE (International Ultraviolet
Explorer) and HST (Hubble Space Telescope) quasars reveal that  there
is a significant steepening of the continuum slope towards wavelength
shorter than $\sim 1000\AA$ (Francis \etal 1991; O'Brien, Gonhalekar
\&  Wilson 1992; Zheng \etal 1997; Telfer \etal 2002). However, for a
limited range in optical and UV, the continuum can be approximated by
a power law $I_{\rm cont}(\nu)\propto \nu^{\anu}$.  The distribution of
indices may not be Gaussian, and may also depend on redshift
(e.g. Telfer \etal 2002).  However, measuring continuum indices
without a very large range of wavelength, or some estimate of the
strength of the contribution from blended emission lines, proves
difficult (e.g. Vanden Berk \etal 2001).  Indeed, Natali \etal (1998) 
have noted that  the value of the continuum index is sensitive to the
precise rest  wavelength regions used for fitting. Therefore, the
steep indices  measured for high-redshift quasars (e.g. Sargent,
Steidel \&  Boksenberg 1989; Schneider, Schmidt \& Gunn 1991; Francis
1996; Fan  \etal 2001) may be due to the restricted wavelength range
used in the  fit, as suggest by Schneider \etal (2001), and not to a
change in the  continuum index with redshift.

\begin{figure}
\resizebox{0.45\textwidth}{!}{\includegraphics{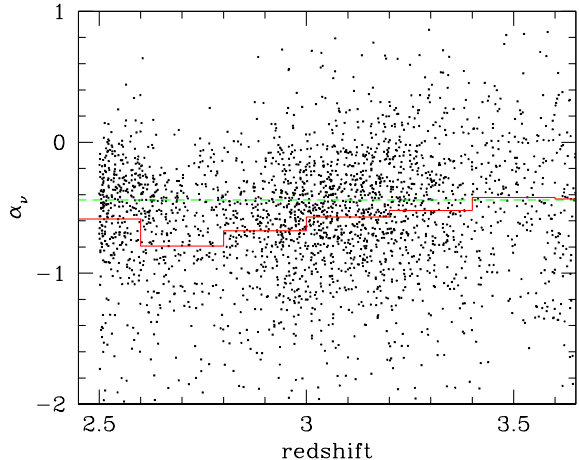}}
\caption{The distribution of continuum indices as a function of 
redshift for quasars with $z\lsim 3.6$. The histogram indicates 
the mean in bins of $\Delta z=0.2$. The averaged continuum index
is $-0.59\pm 0.35$. The horizontal dashed line is $\anu=-0.44$, 
the mean slope reported by Vanden Berk \etal (2001) and B03.}
\label{fig:indices}
\end{figure}

Following P93 and B03, we assume that the shape of the quasar 
continuum is the superposition of a single power law and emission 
lines. Our approach, however, differs from theirs in that we include 
variation in continuum indices. To proceed, we select wavelength 
windows free of emission lines and fit a power law continuum on a 
spectrum-by-spectrum basis.
Although there are essentially no emission-line free regions
(Vanden Berk \etal 2001), we use the rest wavelength intervals 
1450-1470\AA\ and 1975-2000\AA\ (e.g. Telfer \etal 2002). A visual 
inspection of the fitted continuum has convinced us that this 
prescription provides a reasonable description  of the individual 
continua. Note,
however, that low redshift SDSS composites indicate that the window 
1975-2000\AA\ may be contaminated by FeII emission (see Fig.6 of 
Vanden Berk \etal 2001). This would cause us to infer continua 
softer than the actual ones (Telfer \etal 2002). The
distribution of continuum indices is plotted in Fig.~\ref{fig:indices}
for quasars with $z\lsim 3.6$. It is not possible to perform a 
similar measurement of the continuum at higher redshift 
because of the spectroscopic red limit of
9200\AA\ . The histogram indicates the mean value of $\anu$ in bins 
of $\Delta z=0.2$. The average power law slope of the subsample is 
-0.59 (median -0.53) with a 1$\sigma$ dispersion of 0.36. This is in 
good agreement with the mean slope reported by Vanden Berk \etal (2001) 
and B03, $\anu=-0.44$, and with values found in optically selected 
samples (e.g. Francis \etal 1991; Natali \etal 1998).
At $z\bsim 3.6$, fitting a continuum proves 
difficult due to the relatively low signal-to-noise, and the short  
continuum baseline available redward of the op emission line. 
Using the rest wavelength range redward of the
CIV emission line, $\sim$1600-1700\AA\, then plays a significant role 
in determining the slope (Schneider \etal 2001). We have tried to fit 
the regions near 1260\AA\ and 1650\AA\, as done in Schneider \etal (2001). 
As a consistency check, we have remeasured the continuum indices of 
quasars with $z\lsim 3.6$ using that rest frame region. We have found 
that this method gives steeper indices than those inferred from the
rest wavelength region 1460-2000\AA, in agreement with Vanden Berk 
\etal 2001.
We have tried several other alternatives but none of them gave 
satisfactory results. We have therefore opted for a constant slope 
$\anu=-0.44$ when the quasar redshift is $z\bsim 3.6$, implicitly 
assuming that the mean continuum index does not change much with 
redshift. Note that this will affect solely our measurement of the \op 
flux PDF at $z=3.9$.

\begin{figure}
\resizebox{0.45\textwidth}{!}{\includegraphics{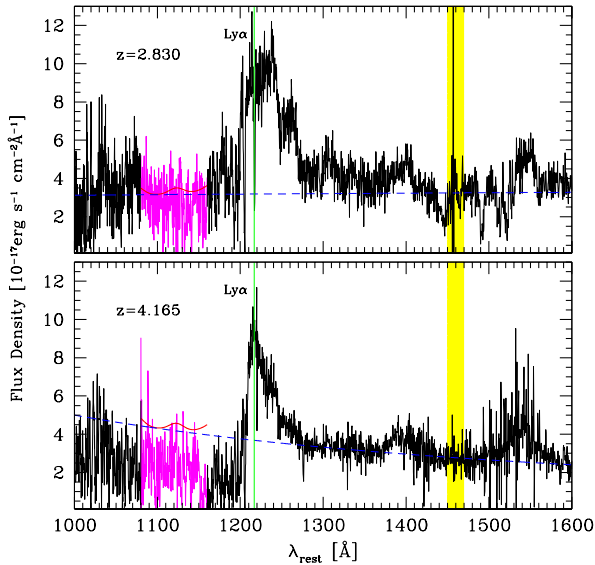}}
\caption{Two quasar spectra of the DR3 sample plotted as a function of
wavelength in the rest frame. The dashed line shows the power law fit.
The continuum index is $\ala=0.10$ and -1.56 for the low and high
redshift spectrum, respectively.  The shaded area indicates the
wavelength range 1450-1470\AA\ used to normalise the  continuum. The
solid curve shows the continuum in the \op forest, including the
emission lines. We analyse the \op forest in the wavelength region
1080-1160\AA.}
\label{fig:spec}
\end{figure}

We follow P93, Zheng \etal (1997) and normalise the QSO continuum in 
the rest-wavelength  range $1450-1470\AA$ to have the same flux as 
the observed spectra. To account for the emission lines blueward of 
1216\AA, we adopt the parametrisation of B03,
\begin{eqnarray}
{\cal C} &=& c_0\left(\frac{\lambda_{\rm
rest}}{\lambda_0}\right)^{\ala}+ c_2\exp\left[-\frac{\left(\lambda_{\rm
rest}-c_3\right)^2}{2c_4^2} \right]\nonumber \\
&&+c_5\exp\left[-\frac{\left(\lambda_{\rm rest}-c_6\right)^2}{2c_7^2}
\right]\nonumber \\ &&+c_8\exp\left[-\frac{\left(\lambda_{\rm
rest}-c_9\right)^2}{2c_{10}^2} \right]\;,
\label{eq:cont}
\end{eqnarray}
where ${\cal C}$ is the continuum as a function of the rest wavelength
$\lambda_{\rm rest}$. The continuum slope $\ala=-\left(2+\anu\right)$
is estimated spectrum-by-spectrum as explained above.
The position of the peak of the \op emission line ($c_9$=1215.67\AA),
and the other two emission lines seen in the composite spectrum
($c_3$=1073\AA\ and $c_6$=1123\AA) are fixed to reduce the  number of
free parameter. The remaining six parameters can be obtained, e.g., 
from a $\chi^2$ minimisation of the difference between the simulated 
and observed composite spectra (B03). Here, we have simply adopted the 
best fit values of B03 that were obtained for a constant power law 
slope $\ala=-1.56$. Two examples of spectra and continuum are shown in 
Fig.~\ref{fig:spec}. The solid curve indicates our fit~(\ref{eq:cont}) 
to the continuum in the \op forest region. Note that the continuum
index of the low redshift spectrum, $\ala=0.10$, differs noticeably 
from the average slope of -1.56.

\subsection{The probability distribution of the transmitted flux}
\label{sub:pdf}

\begin{figure}
\resizebox{0.48\textwidth}{!}{\includegraphics{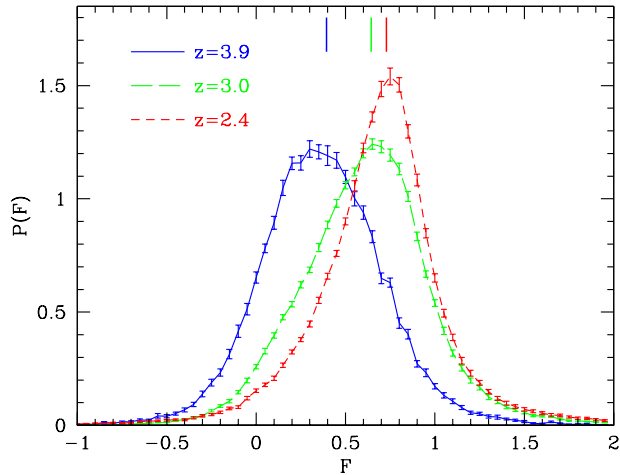}}
\caption{The probability distribution of the
transmitted  flux $F$ is computed from the DR3 sample and plotted in
differential form. Results  are presented for three different redshift
intervals of length  $\Delta z=0.2$ centered at $z=3.9$, 3.0 and
2.4. Vertical bars  indicate the mean transmitted flux $\bar{F}$ in
each redshift interval. The error bars are computed from a jackknife
estimate}
\label{fig:pdss1}
\end{figure}

Once we have taken out the contribution of the continua from our quasar
spectra (on a spectrum-by-spectrum basis), we
compute  $P(F)$, the probability distribution of the transmitted flux
$F$.  Since a substantial fraction of the pixels has a transmitted
flux which lies  outside the interval $0\leq F\leq 1$, we use 60 bins
of width $\Delta F=0.05$ with the first  centered on $F=-1$ and the
last on $F=2$. The first and last bins include the few additional
points with $F<-1$ and $F>2$,  respectively. Results are shown in the
top panel of Fig.~\ref{fig:pdss1}  for three different redshift
intervals as defined in Table~\ref{table1}.  These redshift intervals 
are centered on $z=3.9$, 3.0 and  2.4, allowing a
direct comparison with the high resolution measurements  of McDonald
\etal (2000).  
It is important to note that each redshift interval covers a narrow 
redshift range of $\Delta z=0.2$. This is in order to reduce the
impact of evolution with redshift in each interval.
The vertical bars mark the mean transmitted flux $\bar{F}$ in each
redshift interval, obtained by averaging the individual 
flux pixels (without weighing them according to their noise). 
We have $\bar{F}\simeq 0.39$, 0.64 and 0.73 respectively.
These values are significantly lower than those
inferred from the high  resolution sample, $\bar{F}\simeq 0.45$, 0.69
and  0.82 (M00). The high noise level smoothes severely the PDF
relative to that of high-resolution observations (compare with the
right panels of Fig.~\ref{fig:best}), and is responsible for the
existence of pixels with $F>1$ and $F<0$. The effect is strongest
at $z=3.9$, where the average signal-to-noise in the \op forest is
lowest (see Figure~\ref{fig:pix}).

\begin{figure}
\resizebox{0.45\textwidth}{!}{\includegraphics{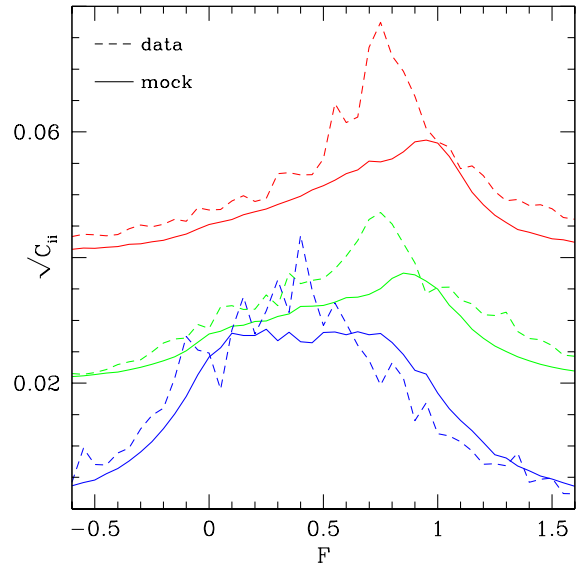}}
\caption{A comparison between the diagonal terms $\sqrt{C_{ii}}$ of 
the covariance matrix computed from the data with a jackknife estimate 
(dashed curve), and the $\sqrt{C_{ii}}$ obtained from a large number
of mock catalogues (solid curves). Results are shown at $z=2.4$, 3 and 
3.9 (from bottom to top). The curves at $z=2.4$ and 3 have been shifted 
vertically for clarity.} 
\label{fig:diag}
\end{figure}

The errors bars attached on the measured PDF shown in
Fig.~\ref{fig:pdss1} are obtained from a jackknife estimate of the
covariance matrix $C_{ij}$, which includes \op forest fluctuations and
measurement noise.  These diagonal elements are plotted as dashed
curves in  Fig.~\ref{fig:diag}. Note that the curves at redshift $z=2.4$
and 3 have  been shifted vertically by 0.04 and 0.02 respectively for
clarity.  As it is difficult to estimate cleanly off-diagonal terms or
diagonal elements lying in the tails of the PDF with such an
estimator, we have also computed the errors from the dispersion
across many realisations of mock spectra with properties similar to
that of  the observed sample (cf. Section~\S\ref{sec:lya}). In
particular, the mock samples have exactly the same total wavelength
coverage (number  of pixels) as the actual sample. The parameters
assume the best fit values inferred in ~\S\ref{sec:param}, with a
mean flux $\la F\ra=0.5$, 0.7 and 0.8. Our estimates of $\sqrt{C_{ii}}$
are shown in Fig.~\ref{fig:diag} as dashed and solid curves, 
respectively. They are consistent with each other at redshift 
$z\bsim 3$. However, at lower redshift, the errors inferred from the 
observed sample are significantly larger than those obtained from 
the mocks. We have found that the jackknife estimator, when applied 
to a single mock SDSS sample, predicts errors similar to those 
inferred from a large number of mock realisations. Therefore, this 
discrepancy probably reflects errors in the measurement of the flux 
PDF, errors in the modelling of low resolution SDSS spectra, or/and
the failure of the lognormal model to adequately describe the low
redshift \op forest (e.g. Nusser \& Haehnelt 1999). Note that the 
off-diagonal terms of
the covariance matrix are negligible presumably because   the large
noise  washes out correlations among the data points. The correlation
coefficient $r_{ij}=C_{ij}/\sqrt{C_{ii}C_{jj}}$ is no larger than
$|r_{ij}|\lsim 0.01$ for $i\ne j$.  This is in contrast with high
signal-to-noise measurements of the PDF (see e.g. M00;  Lidz \etal
2005) where off-diagonal terms in the covariance matrix   are
significant due to the strong correlation among data points.

\begin{table}
\caption{The redshift intervals considered in this paper. The last 
two columns show the number of spectra and pixels included in each 
interval.} 
\vspace{1mm}
\begin{center}
\begin{tabular}{ccccc} 
\hline
$\la z\ra$ & $z_{\rm min}$ & $z_{\rm max}$ & $N$ of spectra & 
$N$ of pixels \\ \hline
2.4 & 2.3 & 2.5 & 942 & 127212 \\
3.0 & 2.9 & 3.1 & 1082 & 133371 \\
3.9 & 3.8 & 4.0 & 281 & 29003 \\
\hline\hline
\end{tabular}
\end{center}
\label{table1}
\end{table}

\begin{figure}
\resizebox{0.48\textwidth}{!}{\includegraphics{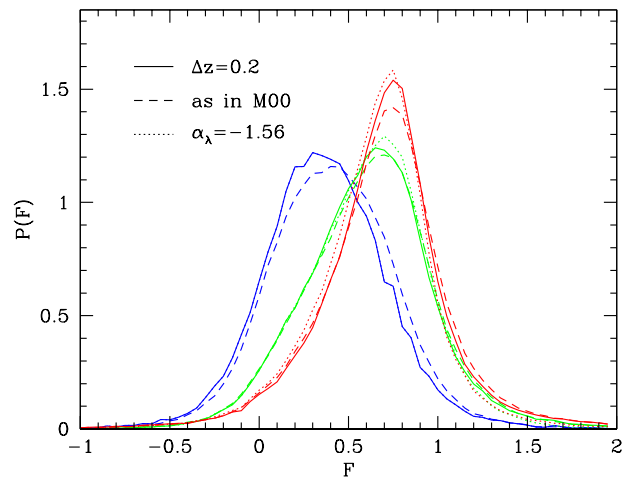}}
\caption{The probability distribution of the transmitted flux for the 
redshift intervals adopted in M00 (dashed curves), for the redshift 
intervals considered in this work (solid curves), and for a fixed 
continuum index $\ala=-1.56$ (dotted curves). The redshift intervals 
are centered at $z=2.4$, 3 and 3.9. }
\label{fig:pdss2}
\end{figure}

\begin{figure*}
\resizebox{0.48\textwidth}{!}{\includegraphics{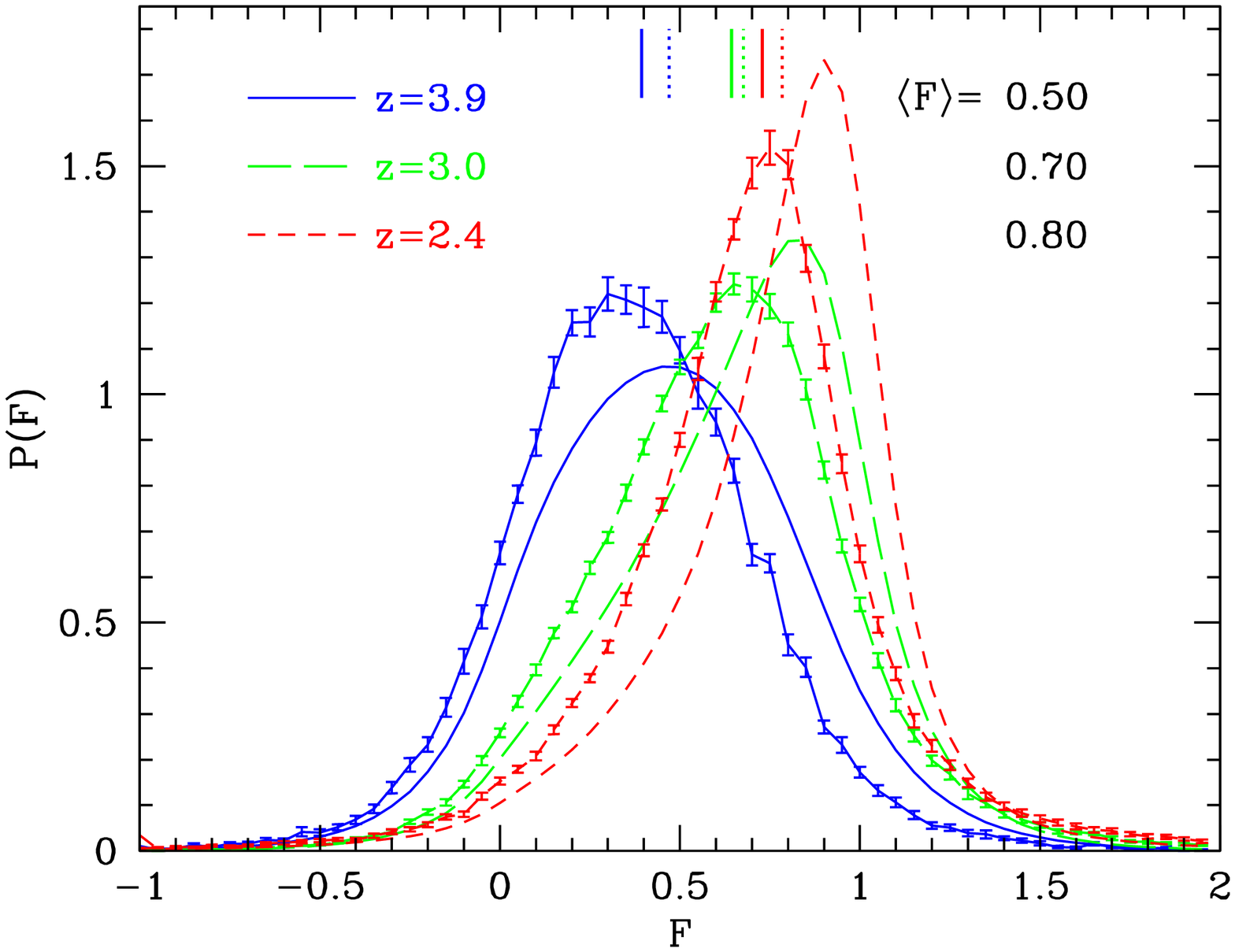}}
\resizebox{0.48\textwidth}{!}{\includegraphics{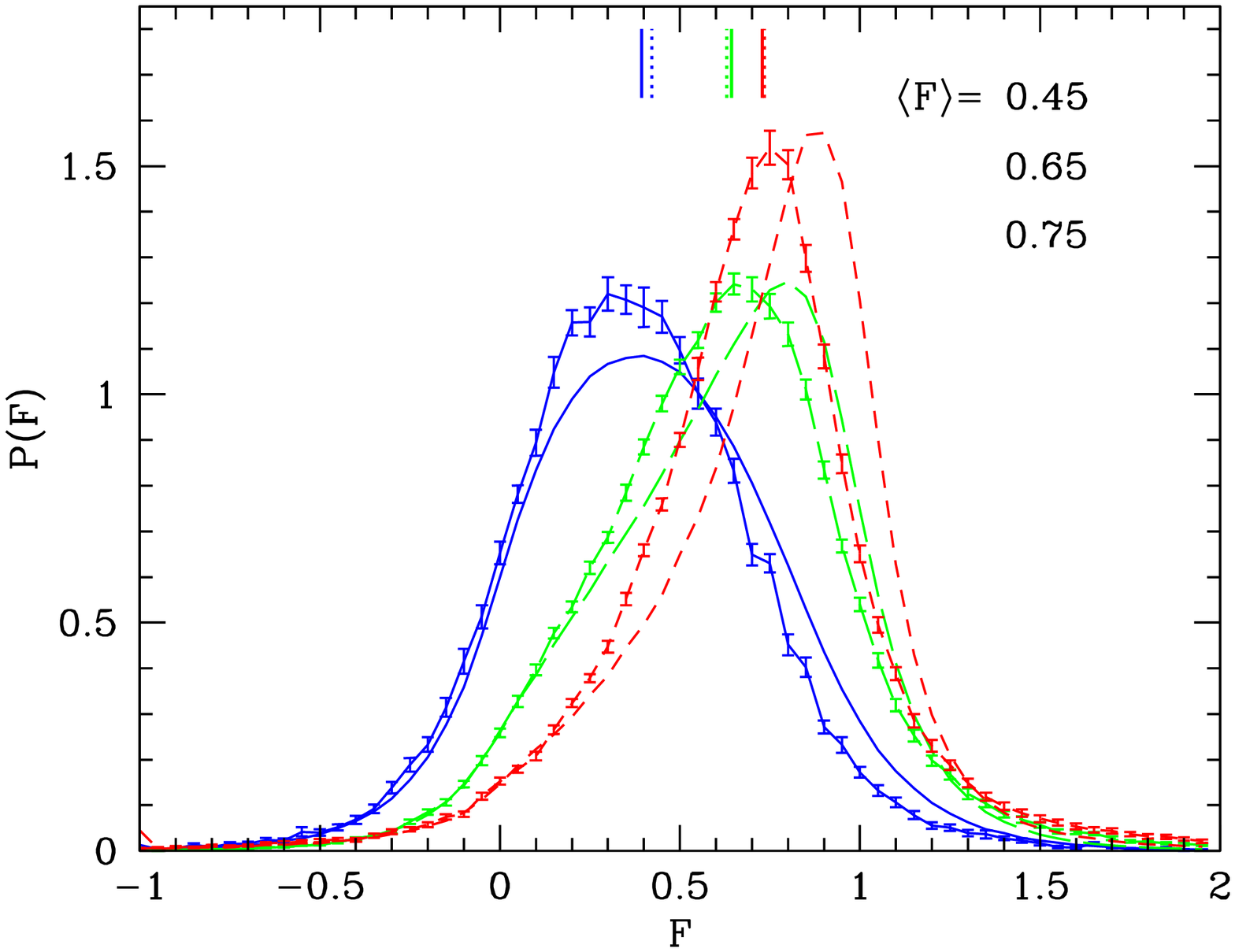}}
\caption{A comparison between the observed and mock \op flux PDFs of
low-resolution QSOs at redshift $z=3.9$, 3 and 2.4. The error bars
are attached to the observed PDF only. The solid and dotted vertical
bars indicate the observed and predicted mean flux $\bar{F}$. In the 
left panel, the models have $\la F\ra=0.50$, 0.70 and 0.80. In the 
right panel, the mean flux parameter is $\la F\ra=0.45$, 0.65 and 
0.75, respectively.}
\label{fig:pdf1}
\end{figure*}

In Section~\S\ref{sec:param}, we have constrained the model parameters
from measurements of the flux PS and PDF which are fairly
representative of the true statistics at redshift $z=2.4$, 3 and 3.9.
However, these measurements are averaged in relatively large redshift
bins, over which the evolution of the \op forest is  significant. It
is, therefore, prudent to examine the extent to which our measurements
of the PDF at redshift $z=2.4$, 3  and 3.9 are sensitive to the
adopted redshift intervals.  We have thus computed the PDF of  
the transmitted flux for the  redshift intervals adopted in M00 
($2.09\leq z\leq 2.67$, $2.67\leq z\leq 3.39$ and $3.39\leq z\leq
4.43$). The
results are shown as dashed curves in Fig.~\ref{fig:pdss2}. They are
compared to our fiducial measurement of the probability distribution 
function obtained with a redshift interval $\Delta z=0.2$ (solid curves). 
The difference is  largest at
$z=3.9$, where  the sharp decrease in the number of pixels at $z\bsim
4$ (Fig.~\ref{fig:pix})  and the strong increase in the flux over the
range $3.5<z<4.5$ conspire to raise the average mean flux by 10 per
cent. At $z=2.4$, the height of the peak is decreased by about 10 per
cent. Obviously, the strength of the effect depends on the exact shape
of the selection function (see Fig.~\ref{fig:pix}).  This result
suggests that the PDF measured by M00 from a sample of  Keck quasars
is also likely to be biased with respect to that  measured in intervals 
of size $\Delta z=0.2$. We will discuss this point later in
Section~\S\ref{sec:discussion}. In Fig.~\ref{fig:pdss2}, the dotted
curves show the PDF at $z=3$ and 2.4 for a fixed value of the
continuum  index, $\ala=-1.56$. Interestingly, accounting for
variation in the  continuum slope has a noticeable impact on the PDF,
especially on  the average mean flux. For a fixed index
$\ala=-1.56$, the mean flux  at $z=3$ and 2.4 is $\bar{F}=0.626$ and
0.675, respectively $\sim$3  and $\sim$7 per  cent lower than the
values of 0.643 and 0.728 obtained with a  spectrum-by-spectrum
fitting. We have also  changed the  interval defining the \op forest,
and found that the  measured PDF is robust to the wavelength range as
long as intrinsic  features to the quasar are excluded.

\section{comparison between observed and mock spectra}
\label{sec:constrain}

In this Section, we compare the flux probability distribution of low
resolution mock spectra with that inferred from the DR3 sample.

\subsection{The PDF of the transmitted flux}
\label{sub:sim}

We generate mock catalogues of low resolution spectra for the
best-fitting values of the parameters obtained in
Section~\S\ref{sec:param}. We adopt a grid similar to that used
in~\S\ref{sec:param}. The  comoving length of a single mock spectrum
is typically  $\bsim 1000\hmpc$. We account for instrumental
resolution, noise and  the presence of strong absorption systems
according to the procedure  outlined in~\S\ref{sub:syn}.

The observed and mock PDFs are compared in the left panel of
Fig.~\ref{fig:pdf1}.  Error bars are attached to the observed PDF
only. The mock spectra have a mean flux parameter $\la F\ra=0.50$,
0.70 and 0.80 at $z=3.9$, 3 and 2.4. This corresponds to an
'effective' (i.e. including strong absorption systems) mean flux
$\bar{F}\approx$ 0.47, 0.67 and 0.78 respectively.  The solid and
dotted vertical bars indicate $\bar{F}$ for the observed and simulated
samples, respectively.  $\bar{F}$ is on average larger by $\bsim 10$
per cent in the mock samples. The mock PDF  correctly accounts for the
shape and redshift evolution measured in the data. However, the 
agreement is poor given the small error bars. At $z\lsim 3$ in 
particular, the peak in the flux probability distribution is 
significantly more pronounced in the mock PDF than in the  
observation.

\subsection{Sensitivity to the mean flux, noise level, and the 
presence of strong absorption systems}
\label{sub:sens}

Since the synthetic spectra have been constrained to reproduce the
observed PDF and PS measured in high resolution data, the 
shortcomings of the lognormal model are unlikely responsible for 
the difference between the observed and mock PDFs.
The disagreement must originate either in the transformation of the 
idealised mocks into realistic looking SDSS spectra, or in the 
measurement of the \op flux probability distribution of the SDSS 
sample. We now discuss a number of systematics that may cause the 
difference between data and simulation. 

\subsubsection{Mean flux}

Fig.~\ref{fig:pdf1} examines the sensitivity of the SDSS PDF to the
mean flux level. The right panel shows the mock PDF of the best-fitting
models listed in Table~\ref{table:best}, with  $\la F\ra=0.45$, 0.65
and 0.75 at $z=3.9$, 3 and 2.4, respectively (This  corresponds to
$\bar{F}=$0.42, 0.63 and 0.74). These values are  marginally
consistent with those inferred from high resolution  measurements of
the \op forest.  Note, however, that the $z\lsim 3$ models poorly
account for the observed \op flux PDF and PS.  Changing $\la F\ra$
affects both the shape and the peak position $F_{\rm peak}$ of the
PDF. The right panel of Fig.~\ref{fig:pdf1}  demonstrates that a
$\sim$10 per cent decrease in $\la F\ra$ improves the agreement with
the  observed PDF, especially with the observed mean flux $\bar{F}$.
Notwithstanding this, at $z\lsim 3$, the mock PDF still peaks at
a higher value  of the transmitted flux $F$ than the observed PDF.
The  effect is strongest at $z\lsim 3$.  Consequently, lowering the
mean flux level $\la F\ra$ in the mock spectra can at best partly
alleviate the tension between the mock and observed PDFs.  It
should also be noted that, in the mock PDF, the mean flux value 
$\bar{F}$ is significantly lower than $F_{\rm peak}$ at $z\lsim 3$. 
This follows from the fact that the PDF is asymmetric around 
$F_{\rm peak}$, decreasing sharply for  $F\bsim F_{\rm peak}$.

\begin{figure}
\resizebox{0.48\textwidth}{!}{\includegraphics{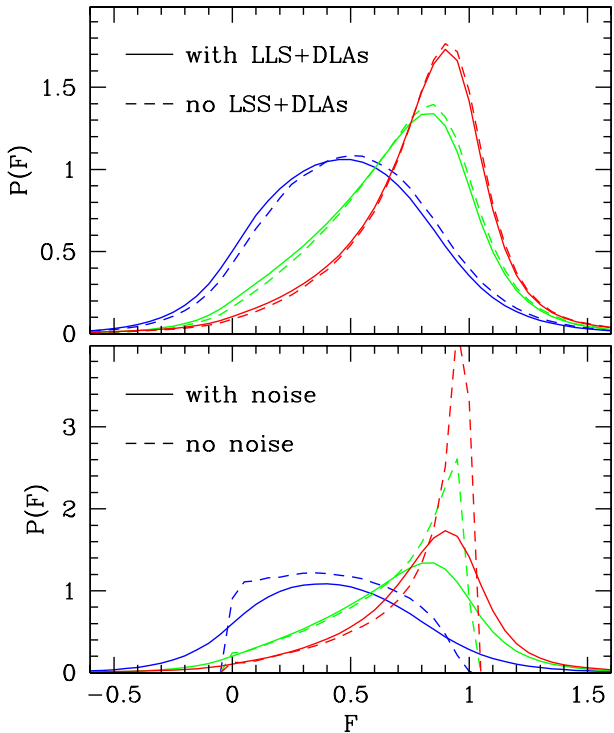}}
\caption{{\it Top panel}~: Impact of strong absorption systems 
on the flux PDF. The solid and dashed curves show the PDF with and 
without strong absorption systems. {\it Bottom panel}~: Sensitivity 
of the PDF to the noise level. The dashed curves show the PDF when 
the noise is set to zero. The mock PDF have $\la F\ra=0.5$, 0.7 and
0.8 for $z=3.9$, 3 and 2.4, respectively.}
\label{fig:pdf2}
\end{figure}

\subsubsection{Metal lines and strong absorption systems}

The probability distribution of the flux may be affected by the
presence of metal  lines and strong absorption systems (e.g.  Schaye
\etal 2003; Viel \etal 2004b).  The top panel of Fig.~\ref{fig:pdf2}
investigates the sensitivity of the PDF to the presence of strong
absorption systems.  The inclusion of strong absorption systems
(cf. Section~\S\ref{sub:syn}) decreases the  mean transmitted flux
$\bar{F}$ of the low resolution mock spectra by $\sim 3-6$ per cent in
the redshift range $2.5\lsim z\lsim 4$.  We confirm the results of
McDonald \etal (2005b) who find that the \op forest is  not very
sensitive to the details of the strong absorption lines,  except when
the damping wings become important.  Regarding the presence of metal
lines,  the typical metallicity of the low-density IGM remains largely
unknown, although early statistical analysis based on pixel optical
depth methods seemed to indicate that there is CIV and OVI associated
with the low-column density \op forest (Cowie \& Songaila 1998;
Ellison  \etal 2000; Schaye \etal 2000a). More recent studies appear
to refute these claims, and suggest that the volume filling fraction
of metals  is small, $\sim 5$ per cent, in the redshift range $2<z<4$
(Pieri  \& Haehnelt 2003; Aracil \etal 2004). Clearly, at  redshift
$z\bsim 3$, absorption in the \op forest region is strongly dominated
by the \op resonant transition, and the impact of metals on the flux
PDF is likely to be negligible. At lower redshift however, metals
contribute more significantly to the absorption. From a sample of
quasars at mean redshift $z=1.9$, Tytler \etal (2004) have estimated
that metal lines absorb $\sim$2.5 per cent of the flux in the \op
forest. However, this falls short of explaining the $\bsim 10$ per 
cent different found between the simulated and predicted PDFs.

\subsubsection{Noise level}

The bottom panel of Fig.~\ref{fig:pdf2} illustrates the sensitivity of
the flux probability distribution to the amount of noise. The solid
curve shows the mock PDF obtained with our default noise level
$\sigma_p$ given by the SDSS reduction pipeline
(cf.~\S\ref{sub:syn}). The dashed curve shows the PDF when  the noise
is set to zero. The large pipeline noise smoothes the idealized,
noise-free PDF  significantly. A few per cent change in $\sigma_p$
noticeably affects the shape of the PDF. In this respect, at $z\lsim
3$, increasing $\sigma_p$ would lower the peak, and  bring the PDF in
better agreement with the observations. Note, however, that changing 
the noise level leaves the mean flux  $\la F\ra$ unchanged.

\begin{figure*}
\resizebox{0.48\textwidth}{!}{\includegraphics{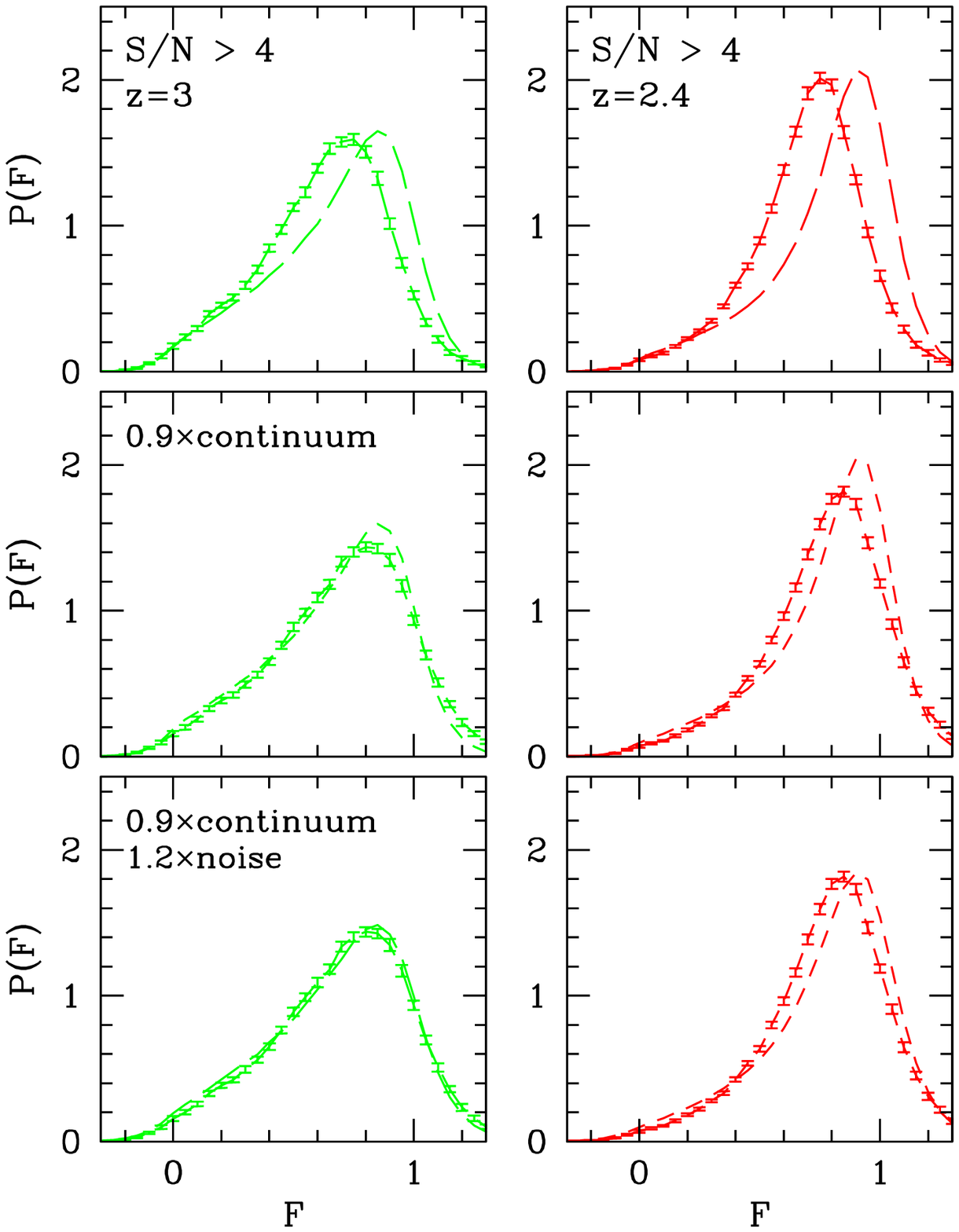}}
\resizebox{0.48\textwidth}{!}{\includegraphics{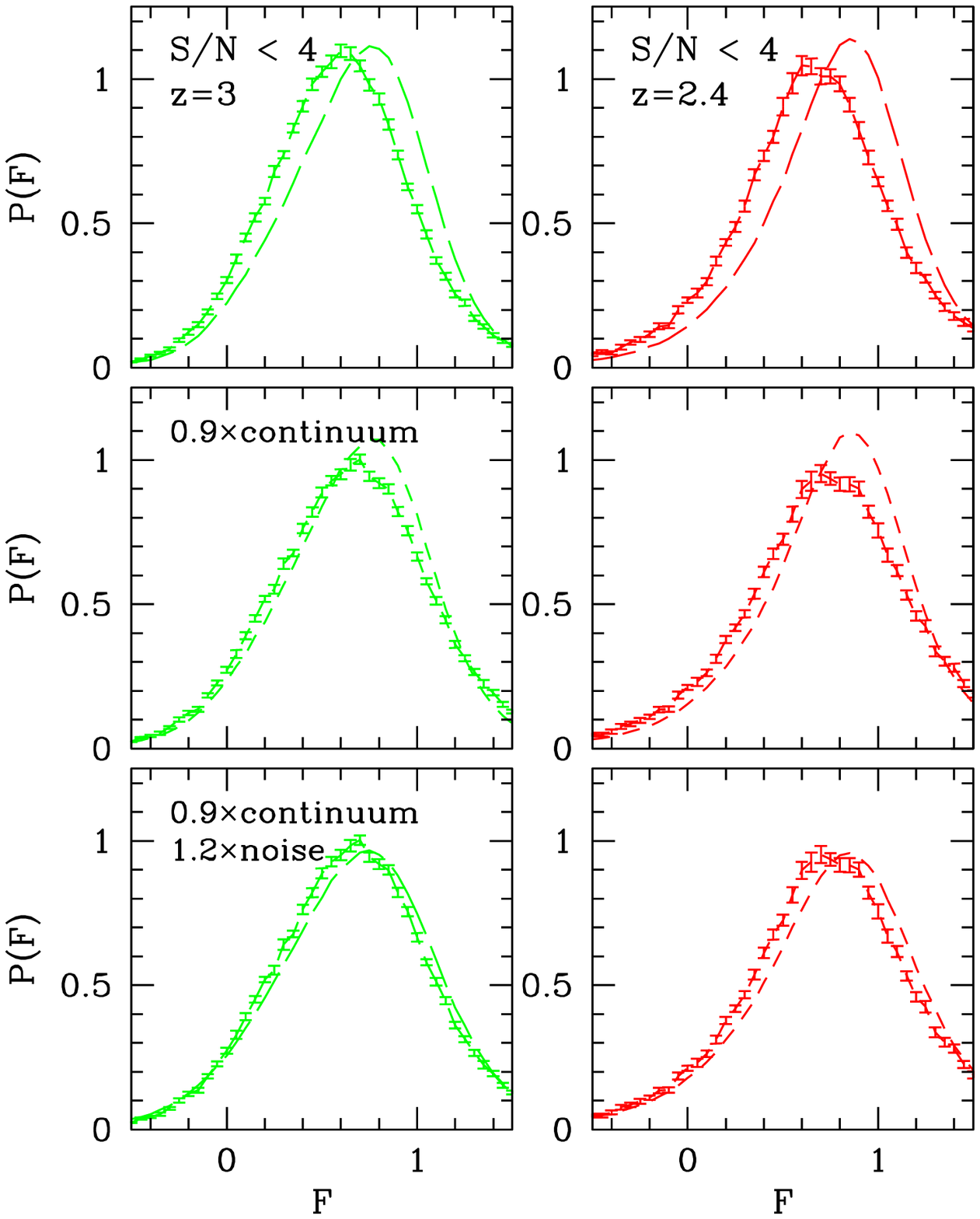}}
\caption{A comparison between the measured (curves with errorbars) and
mock probability distributions of the \op flux at $z=3$ and 2.4. The
left and right  panels are for spectra with $S/N$ ratios greater than
and less than 4,  respectively. The upper panels show results for our
default noise and continuum. In the middle panels, the fiducial
continuum has been  decreased by 10 per cent. In the bottom panels,
both the continuum  and the noise have been rescaled by a factor 0.9
and 1.2, respectively.}
\label{fig:pdf3}
\end{figure*}

Fig.~\ref{fig:pix} shows that the distribution of signal-to-noise in
our sample is broad. To assess the extent to which the shape of the
measured PDF is affected by the noise level, we split the sample based
on the mean signal-to-noise in the \op forest. The upper panels of
Fig.~\ref{fig:pdf3} show the probability distribution of the \op flux
for spectra with $S/N>4$ (left panels) and $S/N<4$ (right panels) for
our fiducial continuum and noise level. Results are shown  solely for
$z=3$ and 2.4, as higher redshift spectra tend to have low  $S/N$
ratios (see Figure~\ref{fig:pix}). The observed PDF is plotted with
errorbars. Cosmic variance errors have been computed separately for
both subsamples, which include   approximately the same number of data
points ($\sim 6\times 10^4$  pixels). The upper  right panels show
that the peak height of the observed flux probability  distribution of
low  $S/N$ spectra is larger at $z=3$ than at $z=2.4$, while high
$S/N$ spectra show the opposite trend. This is presumably due to the
large noise which dominates the signal, and gives the PDF a nearly
Gaussian shape.

Since the SDSS  spectral resolution varies in the range $R\sim
1800-2100$, we  have also computed the PDF with an lower instrumental 
resolution of  150$\kms$ (FWHM). At $z\lsim 3$, a decrease of $\lsim$ 
10 per cent in the instrumental resolution has a relatively small impact  
on the flux probability distribution.

\subsection{Changing the continuum and noise levels}
\label{sub:clevel}

Decreasing the mean flux, accounting for metal lines or increasing
the noise level in the mock spectra can only partly account for 
the difference between the observed and simulated PDF.

In the data, the main sources of systematic errors are inaccuracies
in the continuum fitting of the spectra. Given the large degeneracy
between the amount of absorption and the continuum level in the \op
region, it is unclear whether the single power law approximation can
be extended shortward of the \op emission line.  Low redshift
spectroscopic measurements show indeed that there is a  break around
$1000-1300\AA$ in the slope of the mean quasar  continuum. They
indicate that the continuum turns over in that rest frame region, from
$\anu > -1$ longward of the break to $\anu < -1$  shortward
(e.g. Zheng \etal 1997; Telfer \etal 2002).  The exact location of the
break is, however, difficult to  determine due to the presence of
emission lines.  This turnover is neither accounted for in the
continuum extrapolation  method of P93 and B03, nor in our procedure.
As noticed by  by Kim \etal (2001) and Meiksin, Bryan \& Machacek
(2001), this may lead to an underestimation of $\la F\ra$ that could
be as large as $\lsim 7$ per cent (Seljak, McDonald \& Makarov 2003).
Consequently, we will now relax the assumption of a single power law.
Since there are large  uncertainties in the behaviour of the continuum
blueward of 1216\AA, we have not looked for a parametric form of the
turnover.  Instead, we have simply assumed that the true continuum is
a  rescaled version of our fiducial continuum ${\cal C}$,  ${\cal
C}_{\rm true}=\beta_c\, {\cal C}$, where $\beta_c$ is a mean  correction
factor which we will attempt to constrain.  We ignore any possible
dependence on redshift  and quasar luminosity.

The middle panels of Fig.~\ref{fig:pdf3} show the observed probability
distribution  when the continuum in the \op region is rescaled by 90
per cent  ($\beta_c=0.9$). It is compared to the mock PDF obtained
with the fiducial noise level.  Note that the errors in the observed
transmitted flux depend on the  continuum level, as $F=I_{\rm
obs}/I_{\rm cont}$. The 10 per cent decrease in the continuum
translates into a comparable increase in the observed mean flux, which
is now $\bar{F}=0.69$ and 0.80 at $z=3$ and 2.4,
respectively. Notwithstanding this, the peak  in the simulated PDF is
still more pronounced than in the observed PDF. A further decrease of
the continuum does not improve the  agreement. In fact, unless the
actual mean flux is significantly lower than that inferred from high
resolution data, a substantial increase in the noise level is needed
to reproduce the smooth shape of the observed PDF.

The bottom panel of Fig.~\ref{fig:pdf3} demonstrates that the 
agreement with the data is substantially improved if the pipeline 
noise is increased by $\sim$20 per cent. The agreement is however 
better at $z=3$ than at $z=2.4$, where the mock PDF still appears 
to be shifted to larger values of $F$ when compared to the 
observed PDF. 
This may be due to too large a mean flux level $\la F\ra$. It may 
also reflect the poor performance of the lognormal model at 
redshift $z<3$ (cf. Table~\ref{table:best}). 

The correction factor $\beta_c$ that quantifies the deviation from
a power law is expected to vary from spectrum to spectrum. It also
probably depends on the rest frame wavelength $\lambda_{\rm rest}$. 
Using a constant value of $\beta_c$ smoothes the flux probability 
distribution as compared to the ``true'' PDF, thereby mimicking the 
effect of a larger noise. We have found that the introduction 
of a 10 per cent scatter (Gaussian deviate) in the continuum level 
of $z=3$ mock spectra smoothes the flux PDF on a level comparable to 
a 20 per cent increase in the noise. At $z=2.4$, this corresponds to
an even larger increase in the noise, presumably because the average 
signal-to-noise is lower. Therefore, an increase in the mean noise
level can account for both a larger noise per pixel and variations
in the continuum. We will come back to this point 
in~\S\ref{sec:discussion}.

\subsection{The best-fitting models}
\label{sub:best}

\begin{figure}
\resizebox{0.48\textwidth}{!}{\includegraphics{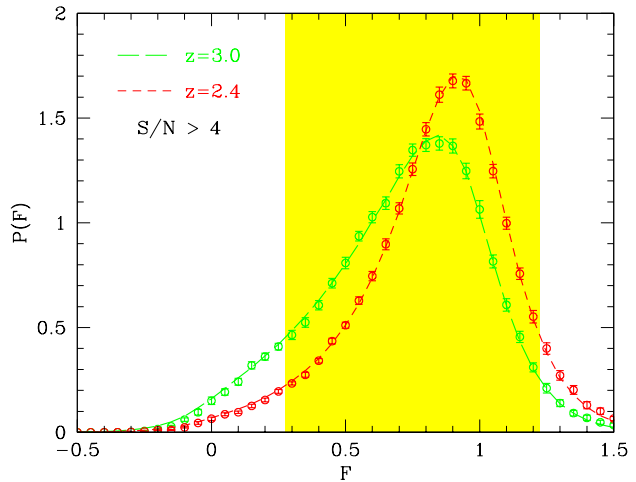}}
\caption{The best-fitting \op flux PDF at redshift $z=3$ and 2.4. The
short and long dashed curves show the best-fitting mock PDF at $z=3$ 
and 2.4 respectively. The continuum level, noise estimate and mean 
flux parameter $\la F\ra$ have been varied to obtain the best-fitting 
models. The symbols with errorbars show the observed flux PDF
of spectra with a signal-to-noise ratio greater than 4. The shaded 
region indicates the SDSS  data points used to compute the value of 
$\chi^2$. We also include in the chi-squared the measurements of
M00 as described in~\S\ref{sec:param}. }
\label{fig:bgt4}
\end{figure}

The agreement between  the mock and observed PDFs of low
resolution spectra can be substantially improved with a simultaneous
decrease in the  continuum level, and an increase in the noise per
pixel. We will now attempt to quantify the correction needed in the
continuum level and noise estimate.  In spite of the large
uncertainties in the actual noise level, we assume  that the true noise
$\sigma_{\rm true}$ differs from the pipeline noise  by a constant
factor, $\sigma_{\rm true}=\beta_n\, \sigma_p$. Similarly, we take the
true continuum in the \op region to be a scaled version of the
fiducial continuum, ${\cal C}_{\rm true}=\beta_c\, {\cal C}$. We let
$\beta_n$ and $\beta_c$ vary in the range $1\leq\beta_n\leq 1.8$ and
$0.8\leq\beta_c\leq 1$. We take $0.65\leq \la F\ra\leq 0.75$ and
$0.75\leq\la F\ra\leq 0.85$ at $z=3$ and 2.4, respectively.  For each
value of $\beta_c$, we compute the PDF of the \op transmitted  flux
from the SDSS quasars with $S/N>4$.  We store the distribution of
noise per pixel  values, $\delta F$, as a function of $\beta_c$ since
the average  $\delta F$ increases with decreasing value of
$\beta_c$. To compute the mock PDF, we let the filtering wavenumber
$\kf$ and the adiabatic  index $\gamma$ assume the best-fitting values
obtained in ~\S\ref{sec:param} as a function of $\la F\ra$, but for a
fixed  value of the temperature, $\thgg=1.5$.  We compute the flux
probability distribution for each choice of  $(\beta_c,\beta_n,\la
F\ra)$.  The goodness of fit of the models is obtained by minimizing 
a $\chi^2$ statistic as in~\S\ref{sec:param}. The shaded region in
Fig.~\ref{fig:bgt4} indicates the SDSS data points we use in 
the calculation of $\chi^2$. We discard the data points falling in 
the $\lsim$10 per cent lower and $\lsim 10$ per cent upper tails
of the flux distribution. We also include in the $\chi^2$ 
minimisation the best-fitting values of $\chi^2$ as a function of 
$\la F\ra$ obtained from the measurements of M00 (cf.~\S\ref{sec:param}). 
The best-fitting values of the parameters are
$(\beta_c,\beta_n,\la F\ra)$=(0.87,1.51,0.72) and (0.84,1.55,0.84),
and correspond to a reduced chi-squared $\chi^2/\nu=1.03$ and 1.54 at
$z=3$ and 2.4 respectively. The best-fitting models are plotted in
Fig.~\ref{fig:bgt4} as short and long dashed curves.
Restricting the $\chi^2$ minimisation to the SDSS data set solely
does not affect noticeably the central values of $\beta_c$, $\beta_n$
and $\la F\ra$.

\section{Discussion}
\label{sec:discussion}

\subsection{Systematics in the measurements}

\begin{figure}
\resizebox{0.48\textwidth}{!}{\includegraphics{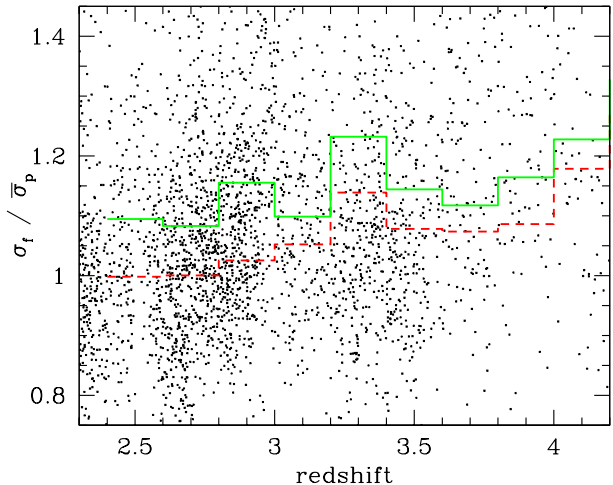}}
\caption{The distribution of ratios $\sigma_f/\bar{\sigma}_p$ in the
rest-frame wavelength interval $1450-1470\AA$, where $\sigma_f$ is the
rms flux  variance, and $\bar{\sigma}_p$ is the mean SDSS pipeline
noise  estimate (see text). The solid and dashed histograms indicate
the  mean and median in bins of $\Delta z=0.2$.}
\label{fig:noise}
\end{figure}

Section~\S\ref{sub:best} argues that the continuum level needs to be
lowered by 10-15 per cent, and the pipeline noise increased by $\sim$ 
50 per cent so that the mock PDF matches the
data. The noise correction is significantly larger than that inferred 
by McDonald \etal (2006) and Burgess (2004) by differencing multiple
exposures of the same quasar. They have found that the
SDSS pipeline underestimates the true errors by 5-10 per cent on
average. Although we do not have access to
additional exposures, we can take advantage of the relative smoothness
of the spectra redward of the \op emission line, and estimate the
noise in, e.g., the rest-frame wavelength interval 1450-1470\AA\ as
the rms variance $\sigma_f$ of  the flux around the
continuum. Then, $\sigma_f$ can be compared to the average pipeline 
noise variance, $\bar{\sigma}_p$, in the same region. The value of 
$\bar{\sigma}_p$ is obtained by
squaring the individual pixel noise estimates,  computing the average
of these squared values, and taking the square root. The distribution
of ratios $\sigma_f/\bar{\sigma}_p$ is shown in  Fig.~\ref{fig:noise}
for the individual spectra. The solid histogram indicates the mean
in bins of $\Delta z=0.2$. The median, which is less  sensitive to
outliers, is also shown as the dashed histogram.  The mean ratio does
not evolve significantly with redshift, although Fig.~\ref{fig:noise}
suggests that it might be bigger at higher  redshift. In the range
$z\sim 2.5-4$, $\sigma_f$ is on average 10-15  per cent larger than
$\bar{\sigma}_p$. This is comparable to the  excess noise contribution
inferred by McDonald \etal (2006) and  Burgess (2004). It is
unclear to which extent  $\sigma_f/\bar{\sigma}_p$ reflects the excess
noise contribution in  the \op forest. However, the fact that McDonald
\etal (2006) find  the same excess noise power in the region
1268-1380\AA\ as they do in the \op forest suggests that the fraction
of extra noise does not depend strongly on $\lambda_{\rm rest}$.
Consequently, the 50 per cent increase in the noise level most 
probably arises from residual variations in the continua of quasars that 
are not accounted for by our spectrum-to-spectrum continuum fitting.
In~\S\ref{sub:clevel}, we have found that the introduction of a 
10 per cent scatter in the continuum level of $z\lsim 3$ mock spectra 
smoothes the flux PDF on a level comparable to a 20-30 per cent 
increase in the noise. 
Hence, we believe that a reasonable $\sim$20 per cent scatter in the 
continuum level can account for the smooth shape of the SDSS PDF if 
the noise excess correction is no larger than $\lsim $10 per cent. 
To proceed further, one could add another free parameter describing 
the residual scatter in the continuum and perform again the $\chi^2$
minimisation of~\S\ref{sub:best}.
However, given our approximate characterisation of the continuum and 
noise level, we have not examined this issue here.

Systematics errors arising from continuum fitting in the measurements 
of M00 may bias our best-fitting values of the parameters, and thereby 
affect the PDF of mock SDSS spectra. However, a comparison with 
hydrodynamical simulations of the
\op forest indicates that the M00 measurements are robust to continuum
errors for transmitted flux values $F\lsim 0.8$.
The large redshift range covered by the M00 bins may also affect 
our results. The latter have been obtained using small redshift 
intervals, $\Delta z=0.2$, to avoid dealing with the (poorly constrained) 
redshift dependence of the model parameters.
Fig.~\ref{fig:pdss2} shows that our measurement of the PDF of SDSS 
quasars is sensitive to the redshift extent of the bins. The 
significance of this effect in measurements of the PDF from 
high-resolution quasars is unknown, though it could be easily estimated 
from the few tens of spectra available so far.

\subsection{Systematics in the model}
 
The lognormal model of the IGM allows us to create very long,
realistic  mock spectra of the \op forest. However, the model  has
several  shortcomings.  It neglects any possible  scatter in the
temperature-density relation of the low density IGM  as a results of
shocks and inhomogeneous helium reionization. It also assumes an
uniform  ultraviolet (UV) background, and a   filtering length that is
independent of  the local gas density and temperature. Furthermore, it
ignores any galactic feedback.  

Hydrodynamical simulations predict
that shock heating should drive a significant fraction of the baryons
into the warm-hot phase of the intergalactic medium (WHIM) at low
redshift.  At the present epoch, this fraction might be as large as
40 per cent (e.g. Cen \& Ostriker 1999; Dav\'e \etal 2001; see also 
Nath \& Silk 2001). At redshift $z\sim 3$ however, these simulations
indicate that this fraction falls below 10 per cent, and that most
of the WHIM baryons resides  in overdensities $\db\bsim 10$
(Dav\'e \etal 2001). Hence, shock heating should have a rather weak
impact on  the low density IGM at $z\approx 3$.  

Inhomogeneities in
the UV background may also affect the power spectrum and the PDF of
the \op flux (Zuo 1992; Fardal \& Shull 1993; Croft \etal 2002a).  At
$z<4$ however, fluctuations due to the finite number of sources  are
only at the few percent level because of the small attenuation  length
(Croft 2004; Meiksin \& White 2004;  McDonald \etal 2005b).  Recent
measurements of the \op absorption near Lyman-break galaxies
(Adelberger \etal 2003) are taken as evidence for the existence of
dilute and highly ionised gas bubbles caused by supernovae-driven
winds. Notwithstanding this, simulations indicate  that their small
filling factor results in a moderate impact on statistics of the \op
forest such as  the power spectrum or the PDF of the  transmitted flux
(e.g. Croft \etal 2002a; Weinberg \etal 2003; Desjacques \etal 2004;
McDonald \etal 2005b; Desjacques, Haehnelt \& Nusser 2006). They may, 
however, have a large impact on the number and properties of absorption 
lines with $\nhi\bsim 10^{16}\cmm$ (Theuns, Mo \& Schaye 2000; Theuns 
\etal 2002a).  

The use of a polytropic
equation of state and a constant filtering length to mimic the
temperature and pressure of the gas  has been shown to produce results
comparable to detailed hydrodynamical simulations (Petitjean \etal
1995; Croft \etal 1998; Gnedin \& Hui 1998; Meiksin \& White
2001). Yet, patchy helium reionization can cause significant
scatter in the temperature-density relation (e.g.  Gleser \etal
2005). Furthermore, in light of the results of Viel, Haehnelt \&
Springel (2006), we expect significant differences between the \op
statistics predicted by full hydrodynamical simulations and from the
lognormal model. Consequently, the constraints on, e.g., the temperature
and adiabatic index which  can be inferred from the high resolution
data should be taken with caution. In particular, the linear amplitude 
$\sgl$ is an effective normalisation which cannot be directly related 
to the actual rms variance of the gas distribution. However, once the 
parameters of the lognormal model are constrained  so as to reproduce 
the observed \op flux power spectrum and probability  distribution 
measured in M00, the disagreement seen in  Fig.~\ref{fig:pdf1} must 
arise either from systematics  errors in the measurement of the PDF 
or in the conversion of the idealised mocks into realistic looking 
SDSS spectra.

\section{Conclusion}
\label{sec:conclusion}

We have presented measurements of the probability distribution of the
\op transmitted flux in the redshift range $2.5\lsim z\lsim 4$, from 
3492 quasars included in the SDSS DR3 data release. 
We have compared the measured PDF to predictions derived from mock 
spectra, whose statistical properties have been constrained to match 
those of high resolution data.
To proceed, we have generated very long, lognormal spectra of the \op
forest that have been degraded to include the instrumental noise and 
resolution of real data. 
The mock spectra provide a good match to the \op flux PS and PDF 
measured in McDonald \etal (2000) in the region $z\bsim 3$.

We have assumed that the quasar continuum follows the parametric form
given in B03. However, unlike B03, we have allowed for the slope of 
the power-law continuum to vary from object to object. We measure 
an average continuum slope of $\anu-0.59\pm 0.36$ in the range 
$2.4\lsim z\lsim 3.6$, in good agreement with the mean slope reported 
by Vanden Berk \etal (2001), $\anu=-0.44$, and with values found 
in optically selected samples (e.g. Francis \etal 1991; Natali \etal 
1998). Accounting for variation in continuum indices has a significant 
impact on the mean flux $\bar{F}$. We find that $\bar{F}$ at redshift
$z=3$ and 2.4 is respectively 3 and 7 per cent higher than the mean
flux measured for a fixed index $\anu=-0.44$.

Although the model parameters have been adjusted to reproduce the
observed \op flux PS and PDF of high resolution data, the mock  SDSS
spectra predict a probability distribution that is significantly  
different from the PDF we measure from the SDSS quasar sample. 
Allowing for a break in the continuum and, more importantly, for 
residual scatter in the continuum level improve the agreement 
substantially. We find that the introduction of a  10 per cent scatter 
in the continuum level of $z\lsim 3$ mock  spectra  smoothes the flux 
PDF on a level comparable to a 20-30 per cent  increase in the noise. 
A combined fit of the SDSS and Keck data indicates that a  decrease of
10-15 per cent in the amplitude of the power law  continuum together
with a 20 per cent scatter can account for the data, provided that the
noise excess correction is no larger than $\lsim$10 per cent.

Measuring the probability distribution of the transmitted flux
requires  a spectrum-by-spectrum treatment of the quasar continuum, as
the latter varies significantly from quasar to quasar. Furthermore, as
we have seen,  it is crucial to account for the slow variation of the
continuum slope in the \op region in order to obtain a sensible
estimate of the flux probability distribution.  Therefore, it would be
desirable to obtain high resolution exposures of a subsample of SDSS
quasars so as to quantify the errors introduced by the continuum
fitting procedure described in this paper. Alternatively,  Lidz \etal
(2005) have suggested working with the estimator defined by
$\delta_f(\omega)=\left(I_{\rm obs}^r(\omega)- I_{\rm
obs}^R(\omega)\right)/I_{\rm obs}^R(\omega)$, where $I_{\rm obs}^r$
and $I_{\rm obs}^R$ is the observed flux smoothed with a Gaussian
filter on scale $r\ll R$ and $R\sim 500\kms$, respectively. They have
demonstrated that the PDF of $\delta_f$ is insensitive to the shape 
and  normalisation of the continuum.
However, this estimator has an important drawback as it
also smoothes out any feature in the redshift evolution of the mean
optical  depth, $\tef(z)$, whose scale is larger than $R$. Their 
choice of $R$ corresponds to a redshift interval $\Delta z\lsim 0.01$ 
at $z=3$. It is therefore unclear whether the sudden change measured 
by B03 at $z\sim 3.2$ can  be detected in the statistics of flux 
estimators other than the transmitted flux $F$.

\section{Acknowledgement}

Special thanks to Scott Burles who made the DR3 sample available, 
and to Joe Hennawi, Joanne Cohn and Martin White for organizing the 
meeting in Berkeley which brought us all together.
We acknowledge stimulating discussions with Mariangela Bernardi,  Ari
Laor, Dan Maoz and Matthew Pieri.  This Research was supported by the
United States-Israel Bi-national Science Foundation (grant \# 2002352).
V.D. thanks the University of Pittsburgh for its kind hospitality, and 
acknowledges the support of a Golda Meir fellowship at the Hebrew
University.

The SDSS is managed by the Astrophysical Research Consortium (ARC) for
the Participating Institutions. The Participating Institutions are The
University of Chicago, the Institute for Advanced Studies, the Japan
Participation Group, The John-Hopkins University, the Korean Scientist
Group, Los Alamos National Laboratory, the Max-Planck Institute for
Astronomy, the Max-Planck Institute for Astrophysics, New Mexico State
University, University of Pittsburgh, University of Portsmouth, 
Princeton University, the United States Naval Observatory and the
University of Washington.

\end{document}